\documentclass[twocolumn,tighten]{aastex7}

\begin{document}

\title{Young Planets around Young Accreting Stars: II. Circumplanetary Disk Formation and Accretion}

\author[gname=Arturo,sname=Cevallos Soto]{Arturo Cevallos Soto}
\affiliation{Department of Physics and Astronomy, University of Nevada, Las Vegas, 4505 S. Maryland Pkwy, Las Vegas, NV, 89154, USA}
\affiliation{Nevada Center for Astrophysics, University of Nevada, Las Vegas, 4505 S. Maryland Pkwy, Las Vegas, NV, 89154, USA}
\email{arturo.cevallossoto@unlv.edu}

\author[gname=Zhaohuan,sname=Zhu]{Zhaohuan Zhu}
\affiliation{Department of Physics and Astronomy, University of Nevada, Las Vegas, 4505 S. Maryland Pkwy, Las Vegas, NV, 89154, USA}
\affiliation{Nevada Center for Astrophysics, University of Nevada, Las Vegas, 4505 S. Maryland Pkwy, Las Vegas, NV, 89154, USA}
\email{zhaohuan.zhu@unlv.edu}

\begin{abstract}

We present global 3-D ideal-MHD simulations of young giant planets in a protoplanetary disk (PPD) undergoing magnetospheric accretion onto its host star. Extending our previous work, we insert three planets with planet-to-star mass ratio \(q = 0.01\) (\(\sim 10\) Jupiter masses) at 0.45, 0.9, and 1.8 times the magnetospheric truncation radius, \(R_T\) , thereby probing the low-density cavity, disk-cavity boundary, and turbulent inner disk.
The outermost two planets develop circumplanetary disks (CPDs), whereas the innermost planet forms no persistent bound structure because its high velocity relative to the magnetosphere reduces the effective Bondi radius below the planet's size.
The outermost planet forms a stable, coplanar CPD whose inner region is rotationally supported and rotates at nearly the local Keplerian speed within \(\sim 0.25\) Hill radii. High-latitude inflow and midplane decretion produce meridional circulation and continuous recycling between the CPD and surrounding disk. Inflow through the Hill sphere is comparable to previous Hill-based hydrodynamic estimates.
The middle planet, located near \(R_T\), develops a persistent polar CPD whose angular-momentum vector is parallel to the disk midplane, demonstrating that coherent circumplanetary structures can form within highly dynamic magnetospheric flows.
Based on both the effective Bondi radius and thermal-mass criterion, we identify conditions under which massive giant planets form CPDs in the inner disk. Estimated planetary inflow accretion rates imply mass-doubling timescales \(\gtrsim 10\) Myr due to the low disk surface density, suggesting that, in the magnetically active zone, these planets undergo only limited late-stage growth before disk dispersal.

\end{abstract}


\section{Introduction}

Planets form and grow within gaseous protoplanetary disks, and sufficiently massive protoplanets can become surrounded by compact, gravitationally bound structures known as circumplanetary disks (CPDs).
A CPD is the rotationally supported, planet-bound disk inside the Hill sphere, the region where the planet's gravity dominates over the tidal field of the star, and is therefore dynamically distinct from the surrounding protoplanetary disk (PPD).
As gas enters the Hill sphere, it typically carries substantial angular momentum with respect to the planet and cannot accrete directly onto the planetary surface.
Instead, it circularizes into a rotationally supported disk or, depending on the thermodynamics, a thicker disk/envelope structure \citep[e.g.,][]{Lubow_1999,DAngelo_2002,Tanigawa_2012,Szulagyi_2016,Fung_2019}.

CPDs are central to giant-planet formation because they regulate the delivery of mass and angular momentum onto the planet, thereby influencing planetary growth, spin, and accretion luminosity \citep[e.g.,][]{Tanigawa_2012,Fung_2019}.
They are also the natural birth environment of regular satellite systems, providing both the reservoir and the dynamical setting for satellite formation \citep[e.g.,][]{Canup_2002,Canup_2006,Ward_2010}.
The existence of prograde, low-inclination, and low-eccentricity regular satellites around the giant planets in the Solar System strongly suggests that rotationally supported CPDs once surrounded Jupiter and Saturn \citep[e.g.,][]{Fung_2019}, but also possibly Uranus and Neptune \citep[see][]{Szulagyi_2018}. 

Recent direct-imaging and millimeter/submillimeter observations have begun to provide candidate CPD detections, most notably in the PDS 70 system, where the accreting protoplanets PDS 70 \textit{b} and \textit{c} reside inside a dust-depleted cavity and show signatures consistent with circumplanetary accretion \citep[e.g.,][]{Keppler_2018,Wagner_2018,Christiaens_2019,Haffert_2019,Isella_2019,Benisty_2021,Fasano_2025}.
More recently, a third accreting protoplanet, PDS 70 \textit{d}, has also been proposed \citep[see][]{Christiaens_2024}.
Beyond direct continuum and H$\alpha$ detections, CPD accretion and planet-induced heating may also produce observable thermal and chemical signatures in the surrounding disk, including localized molecular-line anomalies associated with planet-hosting regions \citep[e.g.,][]{Zhu_2015a,Cleeves_2015,Izquierdo_2026}.

Because CPDs occupy sub-Hill-sphere scales, depend on intrinsically three-dimensional inflow, shocks, and angular-momentum transport, and at the same time extract mass from the surrounding PPD, their accurate study requires global high-resolution 3-D simulations.
Hydrodynamic simulations have established a basic picture of gas capture and CPD formation around gap-opening planets \citep[e.g.,][]{Lubow_1999,DAngelo_2002,Bate_2003,Ayliffe_2009,Tanigawa_2012}.
A recurring result from 3-D studies is that CPDs are fed by high-altitude inflow through meridional circulation, which impacts the CPD surface in a shock, while midplane flow can be partially excluded or even redirected outward \citep[e.g.,][]{Tanigawa_2012,FungChiang_2016,Szulagyi_2016}.
More recent work has emphasized that the circumplanetary structure may transition between a rotationally supported disk and a pressure-supported envelope depending on the thermal treatment and opacity, and that CPD sizes and rotation rates differ substantially between the isothermal and adiabatic limits \citep[e.g.,][]{Szulagyi_2016,Fung_2019,Schulik_2020}.
Magnetic fields may also influence CPD orientation \citep[see][]{Chametla_2026b} and further reshape CPD accretion.
If the planetary field is strong enough (\(\simeq 100\) G), it may truncate the CPD and lead to magnetospheric accretion \citep[e.g.,][]{Sasaki_2010,Zhu_2015a,Batygin_2020}, while magnetically driven outflows may also develop \citep[e.g.,][]{Gressel_2013}.

Most CPD simulations to date have focused on giant planets at Solar-System-like or wider orbital radii, typically around or beyond a few au, where planets are embedded in comparatively standard protoplanetary disks \citep[e.g.,][]{Ayliffe_2009,Tanigawa_2012,Szulagyi_2016,Fung_2019,Schulik_2020,Gardner_2026}.
However, a substantial population of giant planets is observed much closer to their host stars, including hot Jupiters, whose origins remain debated between in-situ formation, disk-driven migration, and high-eccentricity tidal migration \citep[e.g.,][]{Winn_2015,Batygin_2016,Dawson_2018}.
For planets accreting in, or migrating through, the inner disk, any CPD would evolve in a very different environment, with stronger stellar tides, a smaller Hill sphere, hotter gas, and flows shaped by the stellar magnetosphere.
These same effects can also limit satellite formation and survival around close-in giant planets by reducing the stable circumplanetary region and enhancing tidal evolution \citep[e.g.,][]{Barnes_2002,Domingos_2006,Kipping_2009,Kipping_2022,Bolmont_2025}.

This work builds on Paper I, which used high-resolution global 3-D ideal-MHD simulations with magnetospheric accretion to study the orbital evolution and inner stalling of young planets \citep{CevallosSoto2025}. Whereas Paper I focused on the planets' migration, the present paper examines the gas bound to the planets themselves, focusing on the formation, morphology, and angular-momentum structure of CPDs at different distances from the magnetospheric truncation radius. Our goal is to determine under what conditions a rotationally supported CPD can form in the inner, magnetically controlled disk, and whether magnetospheric flows can produce warped or misaligned circumplanetary structures.

In \S\ref{sec:theo}, we introduce the key physical parameters used to characterize the CPDs.
The numerical model is described in \S\ref{sec:numerical}.
In \S\ref{sec:results}, we present the results for each simulated planet and examine its circumplanetary environment.
In \S\ref{sec:disc}, we discuss the conditions for CPD formation, the transport of angular momentum, and comparisons with previous work.
We summarize our findings and conclusions in \S\ref{sec:conclusion}. 

\section{Theoretical Framework}\label{sec:theo}

\subsection{Angular momentum and Stress}

Angular-momentum transport is key to understanding the formation and evolution of circumplanetary structures.
By averaging the angular-momentum equation in the azimuthal direction, \(\hat{\phi}\), and integrating it in the vertical direction, \(\hat{z}\), we obtain
\begin{equation}
\begin{array}{rcl}
\displaystyle
\frac{\partial}{\partial t}\int \langle \rho \delta v_\phi\rangle\,dz
&=&
\displaystyle
-\frac{1}{R^{2}}\frac{\partial}{\partial R}
\left(
R^{2}\int \left(\langle \rho v_{R}\delta v_{\phi}\rangle-\langle B_{R}B_{\phi}\rangle\right)\,dz
\right)
\\[1.2ex]
&&
\displaystyle
-\frac{\dot{M}_{\rm acc}}{2\pi R^{2}}\frac{\partial}{\partial R}\left(R \langle v_\phi \rangle \right)
-(\langle \rho v_{z}\delta v_{\phi}\rangle
\\[1.2ex]
&&
\displaystyle
-\left.\langle B_{z}B_{\phi}\rangle)\right|_{z_{\min}}^{z_{\max}}.
\end{array}
\label{eq:ang-mom-der}
\end{equation}
Here \(\langle n \rangle\) represents the averaging of the quantity \(n\) in the azimuthal direction, and \(\dot{M}_{\rm acc}=2\pi R \int \rho v_R dz\) in a disk region extending from \(z_{\min}\) to \(z_{\max}\).
To separate mean rotation from fluctuations, we define the perturbed azimuthal velocity
\begin{equation}
\delta v_\phi \equiv v_\phi - \langle v_\phi \rangle.
\end{equation}
In ideal MHD, the transport of angular momentum can be expressed through the sum of a Reynolds (hydrodynamic/turbulent) stress and a Maxwell (magnetic) stress. In cylindrical coordinates we define
\begin{equation}
\begin{array}{ll}
     T_{R\phi} &= \rho v_R \delta v_\phi - B_R B_\phi,\\
     T_{z\phi} &= \rho v_z \delta v_\phi - B_z B_\phi,
\end{array}
\label{eq:T_cyl}
\end{equation}
where \(\rho v_i \delta v_\phi\) and \(- B_i B_\phi\) are the Reynolds and Maxwell terms, respectively.

If we normalize the stresses by gas pressure \(P\), we obtain the dimensionless \(\alpha\)-parameter as a measure of angular-momentum transport
\begin{equation}
\begin{array}{rclcl}
\displaystyle \alpha_{R\phi,{\rm Rey}}
&=&
\displaystyle \langle \rho v_R \delta v_\phi\rangle/\langle P\rangle,
&\qquad&
\displaystyle \alpha_{R\phi,{\rm Max}}
=
-\langle B_R B_\phi\rangle/\langle P\rangle,
\\[1.2ex]
\displaystyle \alpha_{z\phi,{\rm Rey}}
&=&
\displaystyle \langle \rho v_z \delta v_\phi\rangle/\langle P\rangle,
&\qquad&
\displaystyle \alpha_{z\phi,{\rm Max}}
=
-\langle B_z B_\phi\rangle/\langle P\rangle.
\end{array}
\label{eq:alpha_components_cyl}
\end{equation}
Since the vertically integrated \(R-\phi\) stress determines the accretion rate in Equation \ref{eq:ang-mom-der}, we can then define the vertically integrated stress parameter
\begin{equation}
\alpha_{\rm int}(R) = \frac{\displaystyle \int \langle T_{R\phi} \rangle\,dz}{\displaystyle \langle \Sigma c_s^2 \rangle},
\label{eq:alpha_int}
\end{equation}
where $c_s \equiv \sqrt{P/\rho}$ is the local isothermal sound speed and $\Sigma(R)=\int\rho\,dz$ is the surface density.

\subsection{Spherical-polar Coordinate System}

The previous stresses and \(\alpha\)-parameters can be defined in similar ways in spherical-polar coordinates
\begin{equation}
    \begin{array}{ll}
         T_{r\phi} &= \rho v_r \delta v_\phi - B_r B_\phi,\\
         T_{\theta\phi} &= \rho v_\theta \delta v_\phi - B_\theta B_\phi.
    \end{array}
\label{eq:T_sph}
\end{equation}
Thus the angular-momentum equation can be reanalyzed and rewritten in spherical-polar coordinates
\begin{equation}
\begin{array}{rcl}
\displaystyle
\frac{\partial \langle \rho \delta v_\phi\rangle}{\partial t}
&=&
\displaystyle
-\frac{1}{r^3}\frac{\partial}{\partial r}\!\left(r^3\langle T_{r\phi}\rangle\right)
-\frac{\langle \rho v_r\rangle}{r}\frac{\partial (r \langle v_\phi \rangle)}{\partial r}
\\[1.2ex]
&&
\displaystyle
-\frac{1}{r\sin^2\theta}\frac{\partial}{\partial \theta}\!\left(\sin^2\theta\langle T_{\theta\phi}\rangle\right)
-\frac{\langle \rho v_\theta\rangle}{r\sin\theta}\frac{\partial(\sin\theta\, \langle v_\phi \rangle)}{\partial \theta},
\end{array}
\label{eq:ang_mom_stress}
\end{equation}
We then rewrite Equation \ref{eq:ang_mom_stress}
\begin{equation}
\frac{\partial \langle \rho \delta v_\phi\rangle}{\partial t}
= m_{r\phi} + \dot m_{r} + m_{\theta\phi} + \dot m_{\theta},
\label{eq:m-terms}
\end{equation}
On the right-hand side, the first term, $m_{r\phi}$, is the radial gradient of the $r$--$\phi$ stress, which represents angular-momentum transport by turbulence near the midplane and mean magnetic fields at higher latitudes, while the second term, $\dot m_r$, is the radial accretion term and represents angular momentum carried by the radial flow.
The third term, $m_{\theta\phi}$, is the $\theta$-gradient of the $\theta$--$\phi$ stress and measures angular-momentum exchange between different latitudes or disk layers, such as wind torque, while the fourth term, $\dot m_\theta$, is the meridional accretion term and represents angular momentum carried by meridional flow, such as circulation, funnel flows, or disk-wind motions.

\subsection{CPD Length Scales}

The structure and dynamics of circumplanetary material are governed by the Hill and Bondi radii, which measure the spatial extent over which the planet's gravity dominates over stellar tides and gas pressure support, respectively.

For a planet of mass $M_p$ on a circular orbit of radius $r_p$ about a star of mass $M_\star$, the Hill radius is
\begin{equation}
R_{H} = r_p\left(\frac{q}{3}\right)^{1/3},
\label{eq:hill}
\end{equation}
where $q \equiv M_p/M_\star$ is the planet-to-star mass ratio.

The Bondi radius is
\begin{equation}
R_B = \frac{G M_p}{c_s^2} = r_p\frac{q}{h_p^2},
\label{eq:bondi_basic}
\end{equation}
where $h_p=H_p/r_p$ is the local disk aspect ratio and $H_p$ is the scale height, with these quantities and $c_s$ evaluated at the planet's orbital radius.

A useful dimensionless quantity is the thermal mass parameter, or the ratio of the planet mass to the local disk thermal mass,
\begin{equation}
    q_{\rm th} = \frac{R_B}{H_p} =\frac{q}{h_p^3}.
\label{eq:qth}
\end{equation}
It measures how strongly the planet's gravity perturbs the surrounding gas relative to the local thermal support.

The relevant outer scale for bound or strongly perturbed circumplanetary gas is approximately the smaller of the Bondi and Hill radii.
Using Equation \ref{eq:qth}, the Hill radius can be written as
\begin{equation}
    R_H = \left(\frac{q_{\rm th}}{3}\right)^{1/3}H_p = \left( \frac{1}{3}\right)^{1/3} q_{\rm th}^{-2/3} R_B.
\label{eq:qth_hill}
\end{equation}
For low-mass planets, \(R_B<R_H\), pressure support limits the planet's gravitational sphere of influence, whereas for sufficiently massive planets, \(R_B>R_H\), the situation reverses. However, gas outside \(R_H\) is no longer securely bound to the planet because stellar tides dominate.
Thus, after a certain planetary mass, the Hill radius becomes the relevant limiting scale for CPD formation and inflow.

Planets with \(q_{\rm th}<1\) are referred to as sub-thermal, and those with \(q_{\rm th}>1\) are super-thermal.
The former are well embedded in the disk and primarily perturb gas within a pressure-supported Bondi sphere, while the latter perturb gas on scales comparable to or larger than the disk scale height and sample more of the vertical density gradient \citep[see, e.g.,][]{Machida_2008,Fung_2019,Choksi_2023}.

\section{Numerical Method}\label{sec:numerical}

The simulation setup for the disk, the star, and the initial conditions closely follows the methodology outlined in Section 3 of \cite{Zhu_2024} and Paper I. For convenience we briefly summarize the setup here.

For our simulations we solve the ideal-MHD equations with Athena++ \citep[][]{Stone_2020} using a higher-order Godunov scheme and constrained transport. We employ a Cartesian mesh with static mesh refinement to resolve both the inner magnetosphere and the disk efficiently.
The computational domain spans $[-64\,R_0, 64\,R_0]$ in each direction and the stellar radius, \(R_{\star}\), is chosen as \(0.1\,R_0\), where $R_0$ is the code length unit.
The root grid has 320 cells in each direction. We have seven levels of mesh-refinement with the finest level covering $[-R_0,R_0]\times[-R_0,R_0]\times[-0.125\,R_0,0.125\,R_0]$. The finest cell size is $\Delta x_{\rm min}=3.125\times10^{-3}\,R_0$, so that for $(H/r)_{R_0}=0.1$ the disk scale height is resolved by $\sim16$-$32$ cells for $0.5\,R_0 \lesssim r \lesssim 8\,R_0$.

\subsection{Disk Setup}

Close to the star, stellar magnetic fields are so strong that the magnetic pressure becomes larger than the accretion ram pressure inside the Alfvén radius $R_A$, defined as
\begin{equation}
    R_A = R_{\star} \left( \frac{B_{\star}^4 R_{\star}^5}{2 G M_{\star} \dot{M}_{\rm acc}^2}  \right)^{1/7} \:{\rm in \: C.G.S.,}
\label{eq:truncation}
\end{equation}
where \(B_{\star}\) is the stellar magnetic field at the equator, and \(R_{\star}\) is the stellar radius. Previous works \citep[see, e.g.,][]{Koenigl_1991,Bessolaz_2008,Kulkarni_2013,Takasao_2022,Zhu_2024,Zhu_2025} suggest that $R_A \sim R_T$, where \(R_T\) is the magnetospheric truncation radius, representing the radius where the disk density changes sharply.
In our simulation, $R_T$ is very close to $R_0$, which is roughly $\sim10$ times larger than \(R_{\star}\).
For the physical scaling adopted in this work, corresponding to a disk around a Sun-like star, $R_0 \simeq 0.05\,\mathrm{au}$ (see \S5).

The initial disk has a midplane density profile as
\begin{equation}
    \rho_d(R,z=0) = \rho_0 \left( \frac{R}{R_0} \right)^p,
\label{eq:den1}
\end{equation}
with $\rho_0 \equiv \rho_d(R=R_0,z=0)=1$. The time unit $1/\Omega(R=R_0)=T_0/2\pi$, in which $T_0$ is the orbital period at $R_0$.

The temperature is set to be constant at each cylindrical radius:
\begin{equation}
    c_s^2(R,z) = c_s^2(R=R_0,z=0)\left( \frac{R}{R_0} \right)^q.
\end{equation}
By choosing $p=-2.25$ and $q=-1/2$ we obtain a disk surface density $\Sigma \propto R^{-1}$. For the scale height $H=c_s/\Omega_K$, we set $(H/R)_{R=R_0}=0.1$.
Although, thermal/irradiated disk models generally give $H/R\sim0.02$-$0.05$ inside $\sim 1$ au for T Tauri parameters \citep[e.g.,][]{dullemond2010,Bitsch_2015},
we intentionally choose a thicker disk to ensure that the instability growth driven by the magnetorotational instability (MRI), as well as funnel and cavity dynamics, are not under-resolved near $R_T$.

We adopt an adiabatic equation of state with $\gamma=1.4$ and rapid thermal relaxation, implemented as in \citet{Zhu_2015}, i.e., the gas in each grid cell cools almost instantaneously, keeping the disk approximately locally isothermal.

The initial vertical structure and azimuthal velocity are
\begin{equation}
    \rho_d(R,z) = \rho_d(R,0) \exp \left[ \frac{GM_\star}{c_s^2} \left( \frac{1}{\sqrt{R^2+z^2}} - \frac{1}{R} \right)\right],
\end{equation}
and
\begin{equation}
    v_{\phi,d}(R,z) = v_{K} \left[
    (p+q)\left(\frac{c_s}{v_{K}}\right)^2
    + 1 + q - \frac{qR}{\sqrt{R^2+z^2}}
    \right]^{1/2},
\end{equation}
To avoid singularities at the origin, the radius in the right-hand side is taken as $r\rightarrow \max(R,R_\star)$.

\subsection{Star Setup}

The stellar gravity is smoothed near the surface and we set a hydrostatic stellar interior with
\begin{equation}
    a_r(r) =
    \left\{
    \begin{array}{ll}
    -\frac{GM_\star}{r^2}\,\frac{(r-R_\star)^2}{(r-R_\star)^2 + r_{\rm sm}^2}, & r>R_\star,\\[6pt]
    0, & r<R_\star.
    \end{array}
    \right.
\end{equation}
with the star's smoothing length set to $r_{\rm sm}=0.1\,R_0$.

The azimuthal velocity is initialized as
\begin{equation}
    v_\phi(r) = \Omega_\star r + v_{\phi,d}\,\exp \left[-\frac{(r-r_{\rm mag})^2}{r_{\rm sm}^2}\right],
\end{equation}
with $\Omega_\star=0$ for the non-rotating star, and $r_{\rm mag}=0.5\,R_0$ for an initial magnetosphere radius. The stellar magnetic field is modeled as a dipole. We adopt a position-dependent density floor to maintain reasonable time-steps in highly magnetized regions. The field strength and density floor values can be found in \cite{Zhu_2024}.

\subsection{Planet Setup}

After initialization, turbulence is generated from MRI and the magnetic breaking, allowing the flow to relax into a quasi-steady accretion state with sustained variability. Planets are introduced only after this state is established, at $t=65$ orbits (measured at $r=R_0$).

Planets are represented as orbiting, softened point-mass potentials with planet-to-star mass ratio \(q\).
We adopt a gravitational smoothing length \(R_s=0.012\,R_0\), which with \(R_\star=0.1\,R_0\) corresponds to roughly 1.2 Jupiter radii for a Sun-like star.
A smoothing length spanning several grid cells is typically used to prevent the numerical divergence of a point-mass potential; in our case \(R_s\) is comparable to the physical radius of a massive young planet. This allows us to capture the flow structure close to our planets with good fidelity.

All inserted planets have the same mass, $q=0.01$ (equivalent to $10\,M_J$ for a Solar-mass star), and are placed at $r_p = 0.45\,R_0$, $0.9\,R_0$, and $1.8\,R_0$, following a fixed circular orbit at the local Keplerian angular frequency.
These three locations probe distinct environments: the innermost planet orbits inside the low-density magnetospheric cavity, the middle planet resides near the truncation radius where the disk inner edge rises sharply and filamentary
accretion streams are active, and the outermost planet remains well embedded in the dense protoplanetary disk.
This setup enables a controlled comparison of CPD formation and evolution for the same planet mass under different ambient conditions.

\section{Results}\label{sec:results}

\begin{figure*}[t!]
\centering
\includegraphics[trim=0mm 0mm 0mm 0mm, clip, width=7.5in]{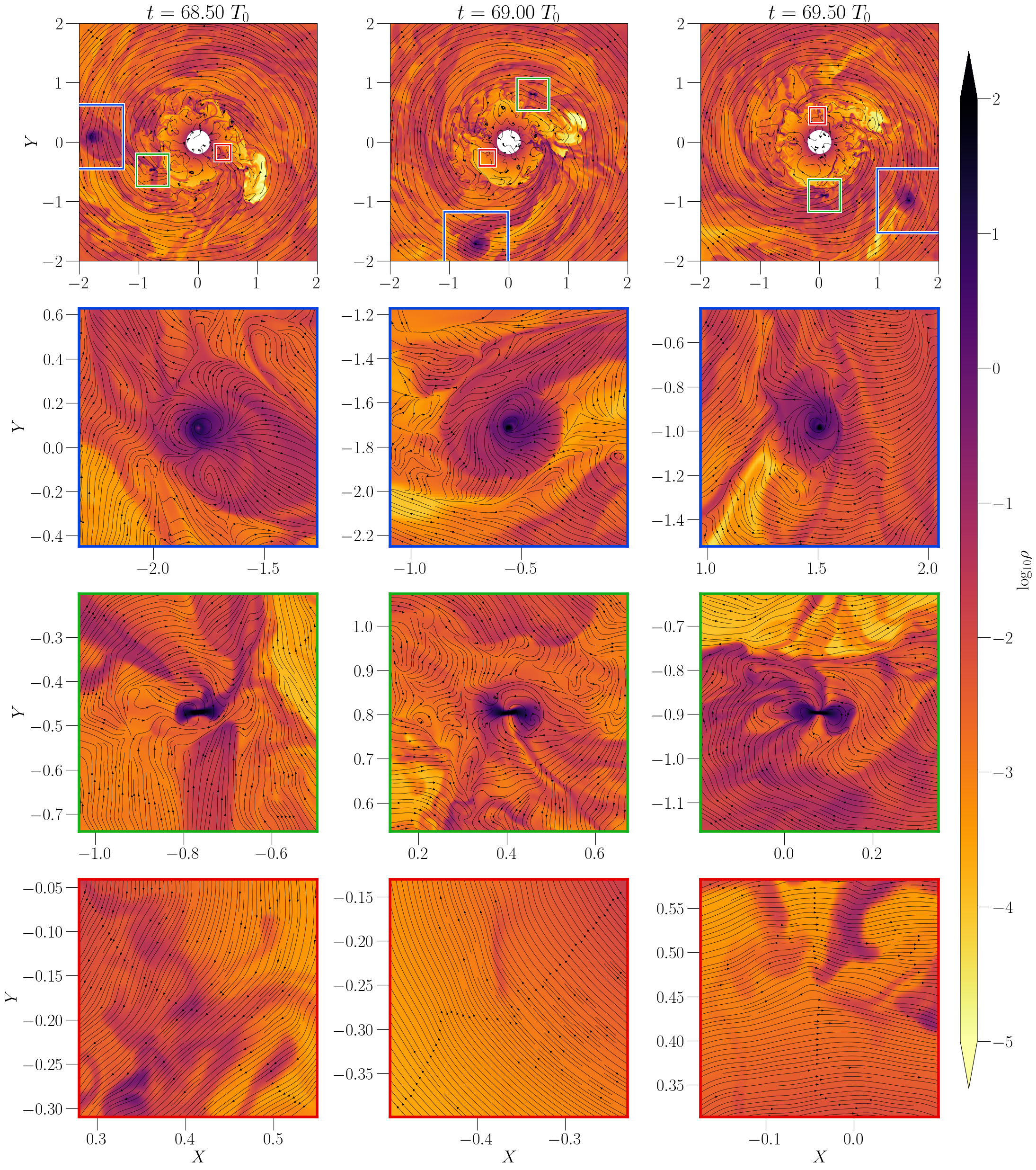}
\caption{
Midplane gas density and velocity streamlines of the simulated protoplanetary disk at three representative times.
\textbf{Top row:} star-centered frame, with the planets marked by colored squares. Each square is centered on a planet and has side length \(4\,R_H\), where \(R_H\) is that planet's Hill radius.
\textbf{Lower rows:} zoom-ins of the boxed regions showing the local gas density and velocity streamlines near each planet. In these panels, the velocity field is shown in the corresponding planet-centered, corotating frame.
}
\label{fig:three_planets}
\end{figure*}

\begin{figure}[t!]
\centering
\includegraphics[trim=0mm 0mm 0mm 0mm, clip, width=\columnwidth]{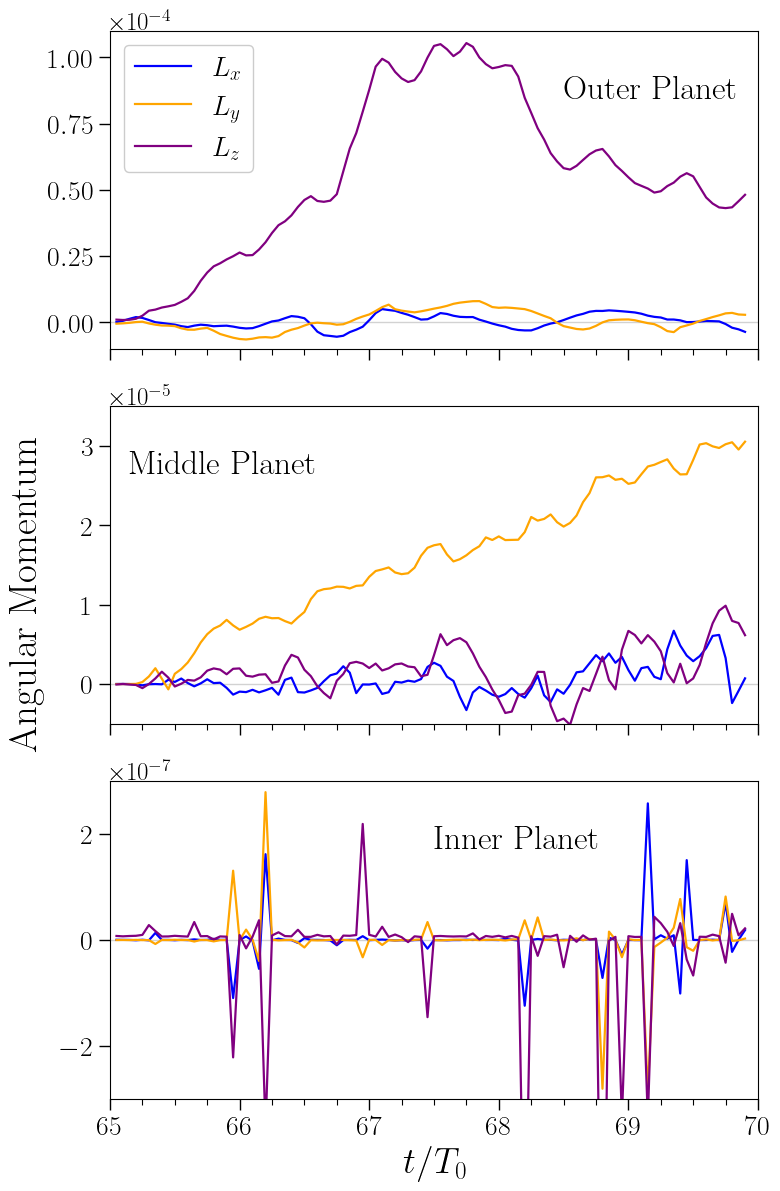}
\caption{
Time evolution of the gas angular-momentum components in the vicinity of each planet.
For each planet, we compute the total angular momentum $\mathbf{L}=(L_x,L_y,L_z)$ of the gas contained in a planet-centered Cartesian box extending $\pm0.6\,R_H$ along each coordinate axis.
Cells located within a spherical region of radius $0.1\,R_H$ around the planet are excluded from the calculation.
}
\label{fig:ang_mom}
\end{figure}

\begin{figure*}[t!]
\centering
\includegraphics[trim=0mm 0mm 0mm 0mm, clip, width=7.5in]{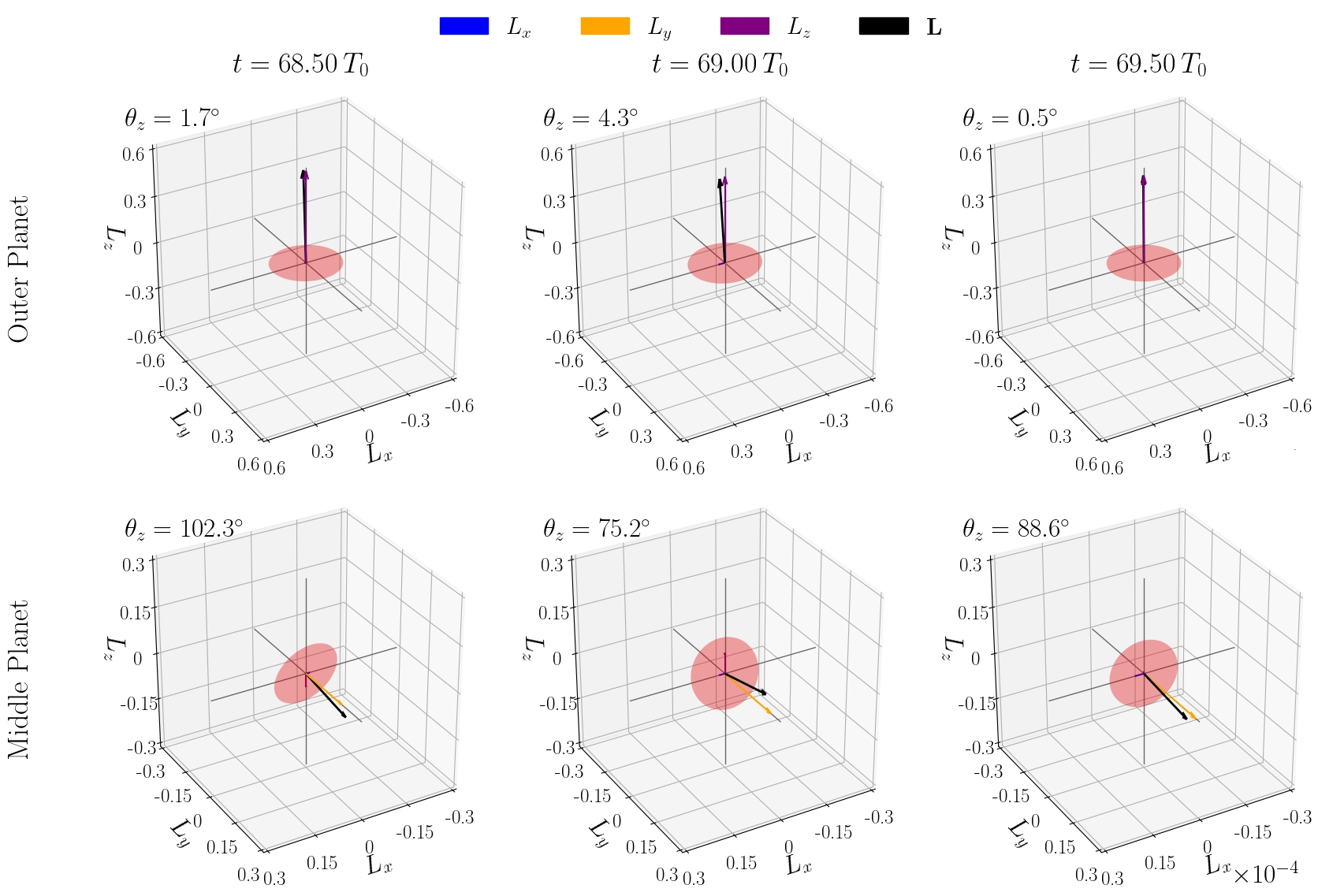}
\caption{
3-D visualization of the total gas angular-momentum vector $\mathbf{L}$ and its Cartesian components for the \textit{outer} (top row) and \textit{middle} (bottom row) planets at the same times shown in Figure \ref{fig:three_planets}.
The semi-transparent red disk is oriented perpendicular to $\mathbf{L}$ and illustrates the approximate instantaneous orientation of the net rotation plane of the circumplanetary gas.
The angle between $\mathbf{L}$ and the $z$-axis, $\theta_z$, is indicated in each panel.}
\label{fig:ang_mom_3D}
\end{figure*}

\begin{figure*}[t!]
\centering
\includegraphics[trim=0mm 0mm 0mm 0mm, clip, width=7.0in]{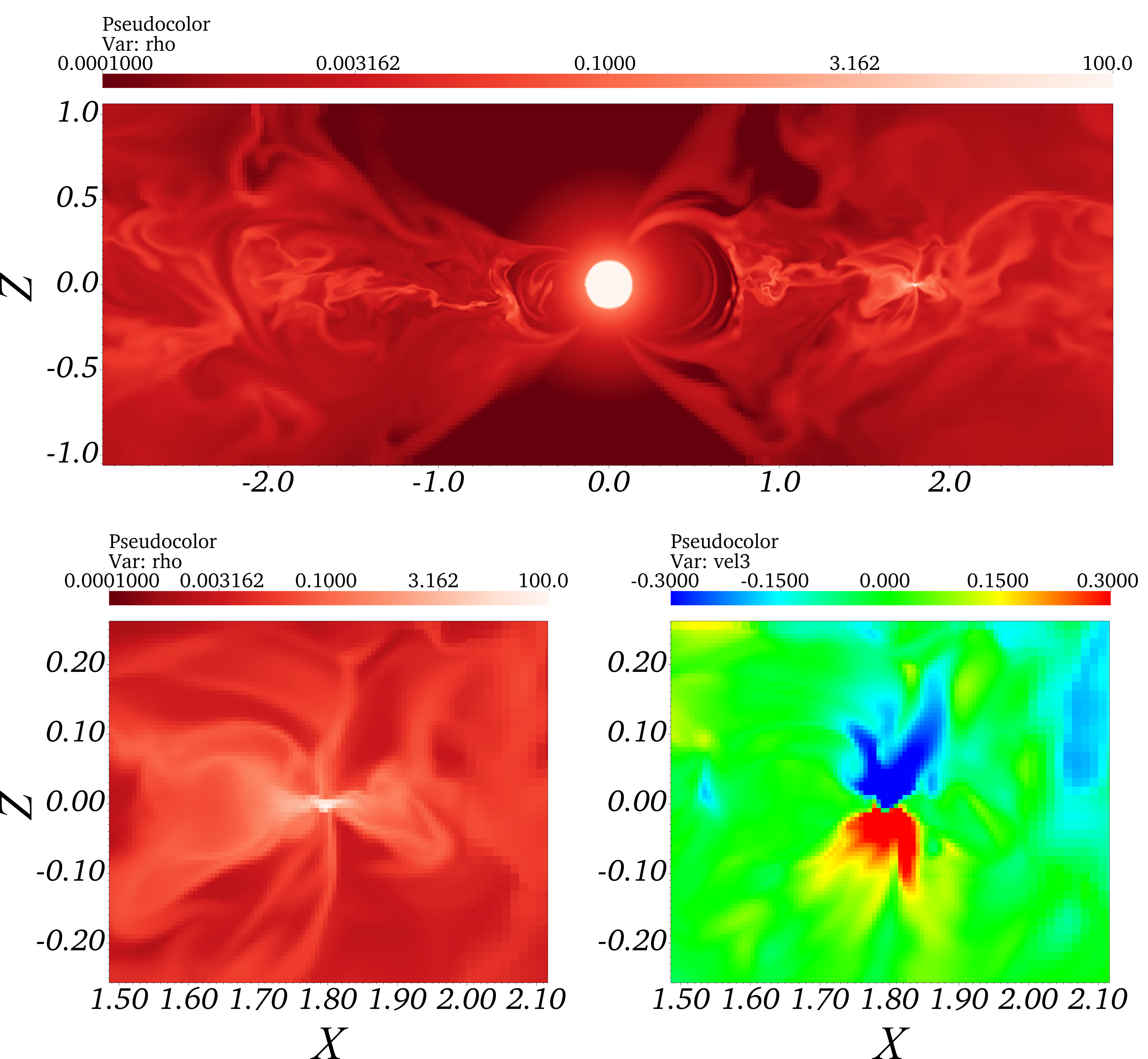}
\caption{
Instantaneous Cartesian \(x\)--\(z\) slices through the \textit{outer planet} at \(t=69.65\,T_0\), taken at \(y=-0.096\). The top panel shows the gas density, \(\rho\), across the PPD, with the star at the origin and the planet at \(x\approx1.8\,R_0\). The lower panels show zoomed views of the planet and its circumplanetary environment in gas density (left) and vertical velocity, \(v_z\) (right). These visualizations were produced using \texttt{VisIt 3.4.2.}
}
\label{fig:slice_p2}
\end{figure*}

\begin{figure*}[t!]
\centering
\includegraphics[trim=0mm 0mm 0mm 0mm, clip, width=7.5in]{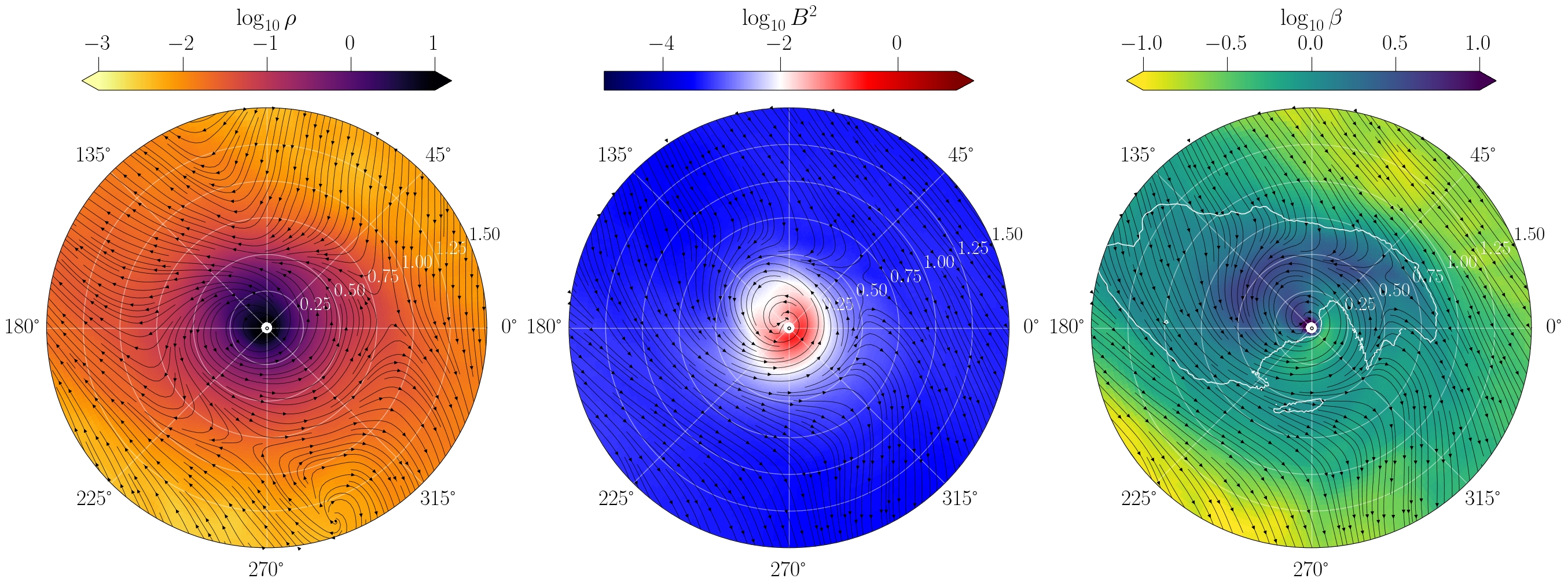}
\caption{
Time-averaged, rotated planet-centered midplane ($\theta=\pi/2$) maps for the \textit{outer planet}. Each snapshot is rotated so that the star lies at a fixed azimuth to the west of the planet. The \(r-\)coordinate is in units of \(R_H\).
From left to right, the maps show the gas density $\log_{10} \rho$, the magnetic energy $\log_{10} B^2$, and the plasma beta $\log_{10} \beta$. The white contour in the plasma beta map marks $\beta=1$. Streamlines in the density map trace the mean in-plane velocity field in the planet-centered, corotating frame, while streamlines in the remaining panels trace the mean in-plane magnetic field.
}
\label{fig:out_midplane}
\end{figure*}

\begin{figure*}[t!]
\centering
\includegraphics[trim=0mm 0mm 0mm 0mm, clip, width=7.5in]{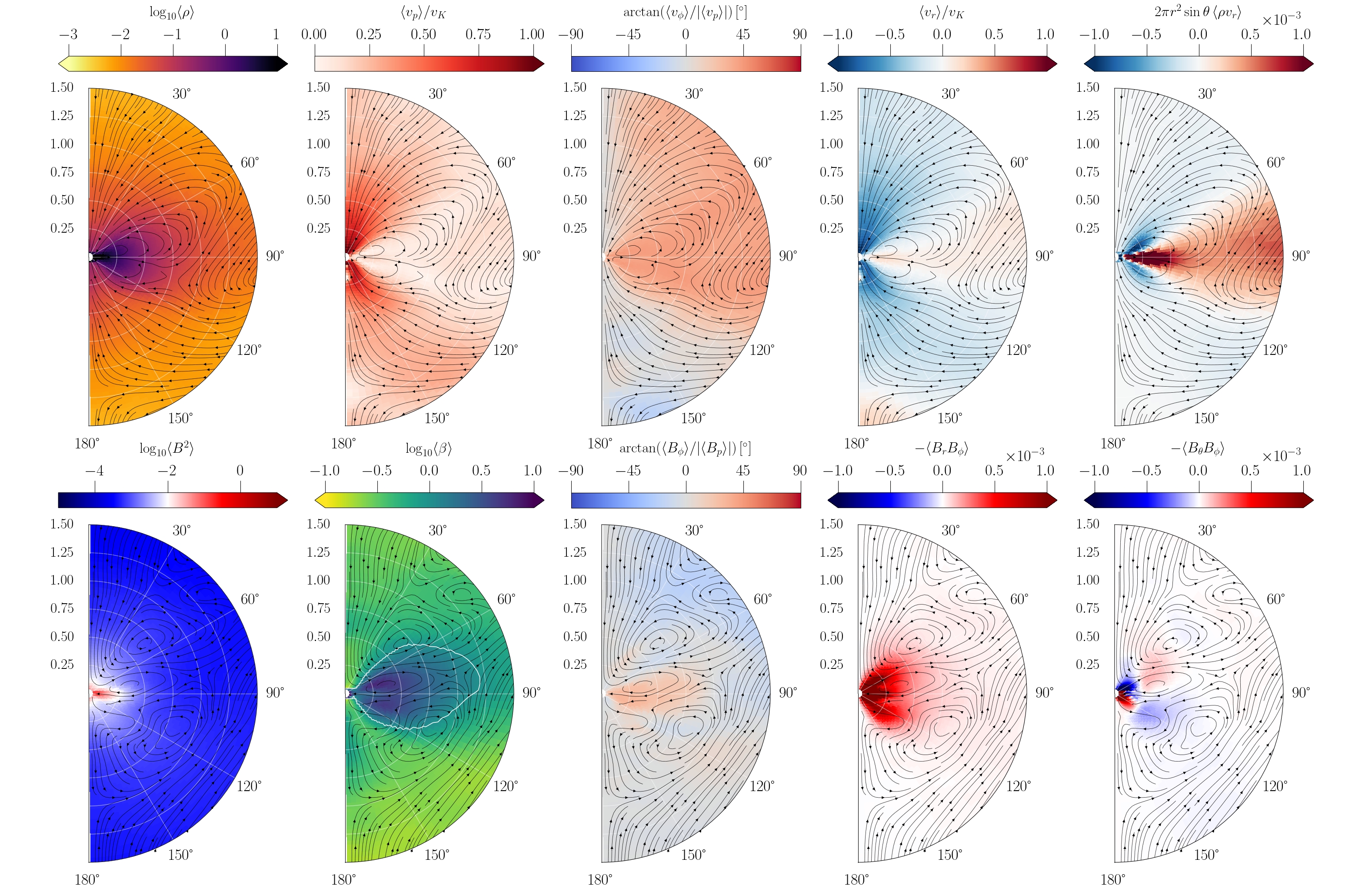}
\caption{Time- and azimuthally-averaged poloidal maps in the planet-centered frame of the \textit{outermost planet}. The \(r-\)coordinate is in units of \(R_H\).
The upper panels (left to right) show the gas density $\log_{10} \langle \rho \rangle$, the poloidal speed normalized by the local planet-centered Keplerian speed $\langle v_p\rangle/v_K$, the velocity angle $\arctan(\langle v_\phi\rangle/|\langle v_p\rangle|)$ in degrees, the radial velocity normalized by the Keplerian speed $\langle v_r\rangle/v_K$, and the mass flux $2\pi r^2\sin\theta\,\langle \rho v_r\rangle$. The lower panels (left to right) show the magnetic energy $\log_{10} \langle B^2 \rangle$, the plasma beta $\log_{10}\langle \beta \rangle$, the magnetic-field angle $\arctan(\langle B_\phi\rangle/|\langle B_p\rangle|)$ in degrees, and the Maxwell stress components $-\langle B_rB_\phi\rangle$ and $-\langle B_\theta B_\phi\rangle$. The white contour in the $\log_{10}\beta$ panel marks $\beta=1$. Streamlines in the upper panels trace the mean poloidal velocity field, while those in the lower panels trace the mean poloidal magnetic field.}
\label{fig:out_poloidal}
\end{figure*}

\begin{figure*}[t!]
\centering
\includegraphics[trim=0mm 0mm 0mm 0mm, clip, width=7.in]{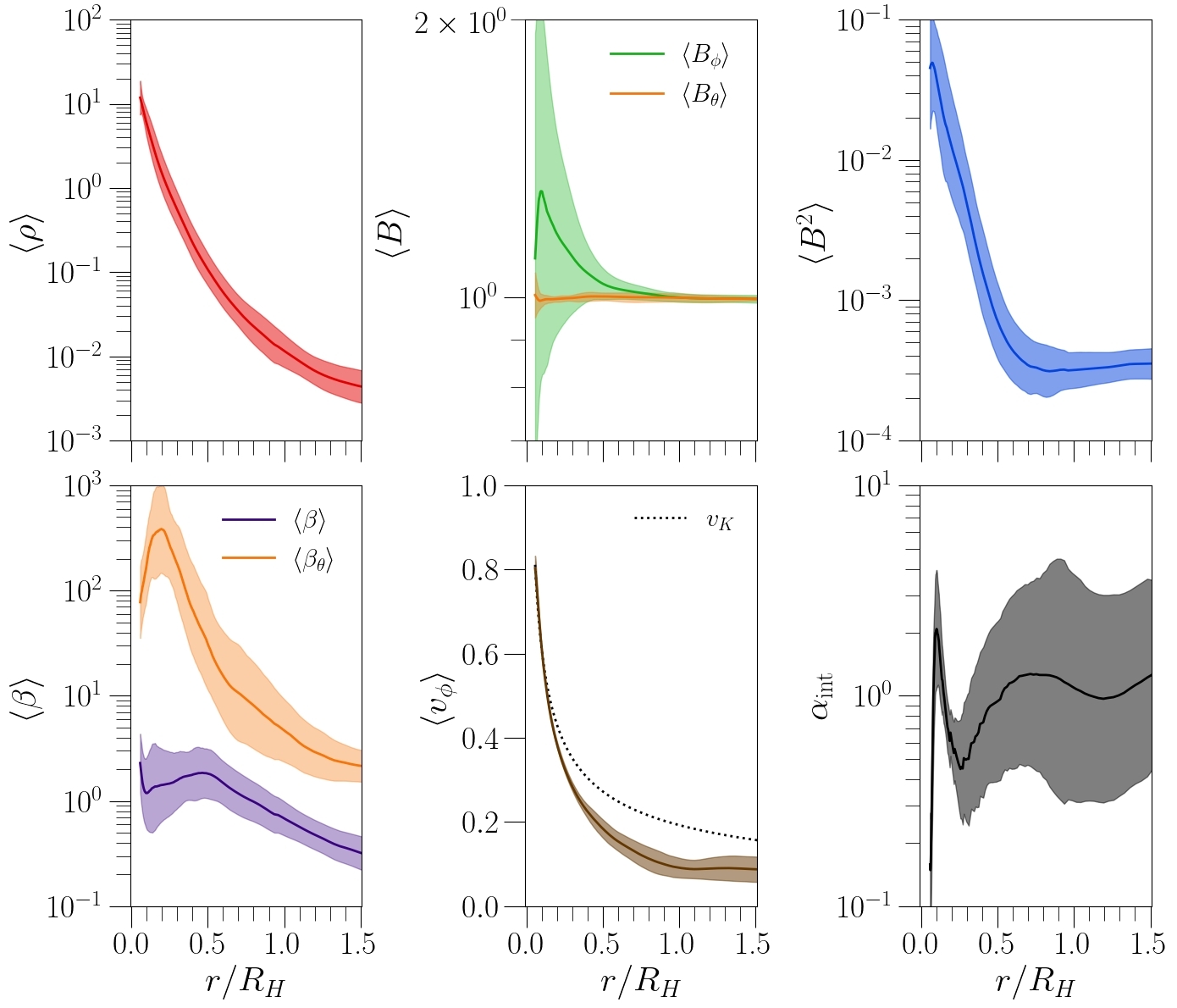}
\caption{
One-dimensional radial cuts for the \textit{outer planet} in the planet-centered frame, shown as functions of distance from the planet in units of the Hill radius, $R_H$. All profiles are azimuthally averaged at each snapshot and then time averaged the final \(2.5\,T_0\) interval. Solid lines show the mean profiles and shaded regions indicate the temporal scatter. From top left to bottom right, the panels show: the gas density $\rho$; the magnetic-field components $B_{\phi}$ and $B_{\theta}$; the magnetic energy density $B^{2}$; the plasma beta $\beta$ and its $\theta$-component, $\beta_{\theta}$; the azimuthal flow velocity $v_{\phi}$ together with the local Keplerian velocity $v_{K}$ (dotted line); and the vertically integrated alpha, $\alpha_{\mathrm{int}}$.
}
\label{fig:out_linear}
\end{figure*}

\begin{figure*}[t!]
\centering
\includegraphics[trim=0mm 0mm 0mm 0mm, clip, width=7.0in]{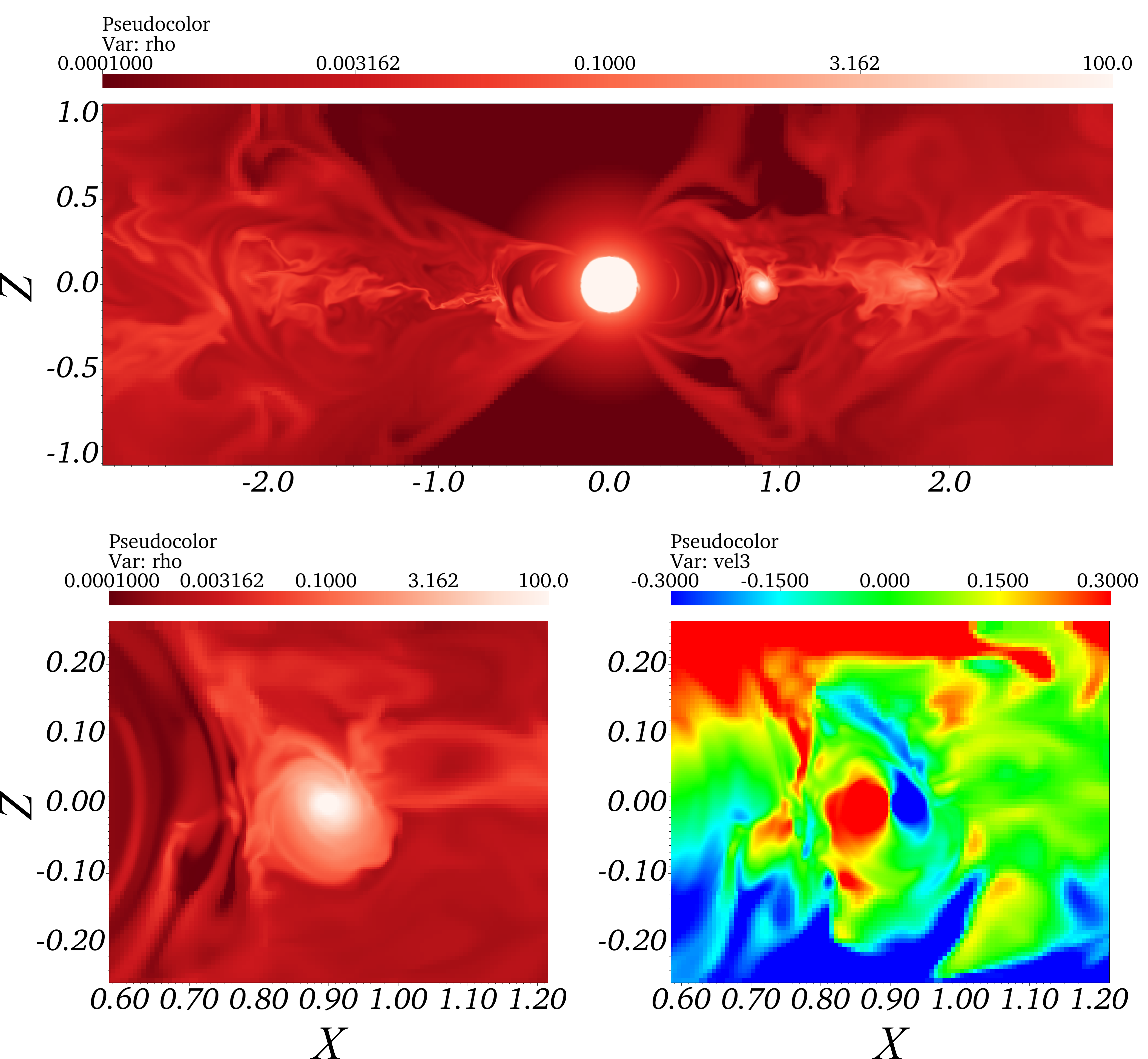}
\caption{
Instantaneous Cartesian \(x\)--\(z\) slices through the \textit{middle planet} at \(t=69.65\,T_0\), taken at \(y=-0.005\). The top panel shows the gas density, \(\rho\), with the star at the origin and the planet at \(x \approx 0.9\,R_0\). The lower panels show zoomed views of the planet and its circumplanetary environment in gas density (left) and vertical velocity, \(v_z\) (right). These visualizations were produced using \texttt{VisIt 3.4.2.}
}
\label{fig:slice_p1}
\end{figure*}

\begin{figure*}[t!]
\centering
\includegraphics[trim=0mm 0mm 0mm 0mm, clip, width=7.5in]{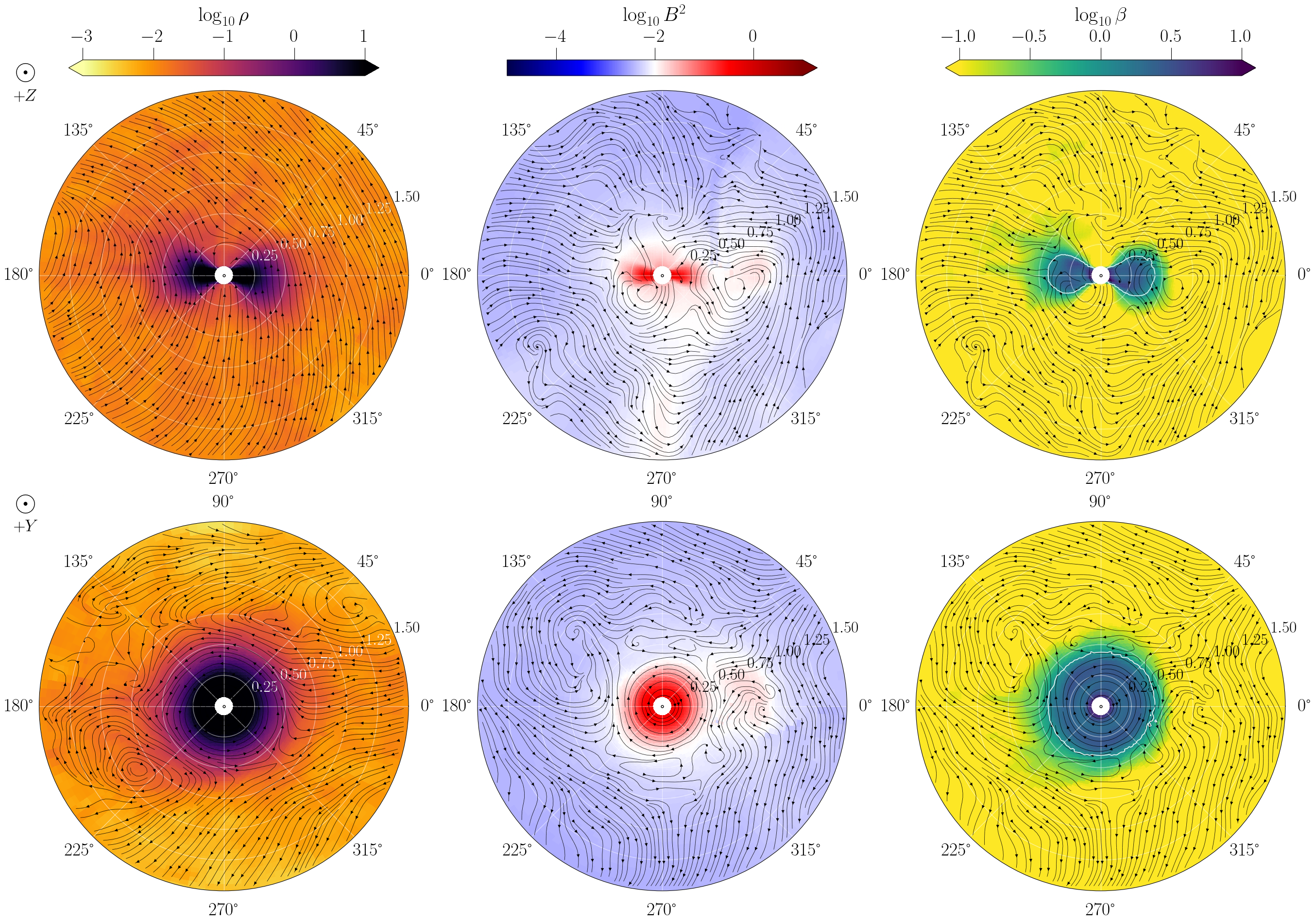}
\caption{
Time-averaged, planet-centered maps for the \textit{middle planet}. \textbf{Top row:} orbital-midplane cut ($\theta=\pi/2$, with $+\hat{z}$ pointing out of the page), in which the circumplanetary disk appears approximately edge-on. \textbf{Bottom row:} perpendicular cut (with $+\hat{y}$ pointing out of the page), revealing the disk approximately face-on.
Unlike Figure \ref{fig:out_midplane}, snapshots are not azimuthally rotated prior to averaging. From left to right, the maps show the gas density $\log_{10}\rho$, squared magnetic-field strength $\log_{10}B^2$, and plasma beta $\log_{10}\beta$. The white contour in the $\log_{10}\beta$ panel marks $\beta=1$. Streamlines in the density panels trace the mean in-plane velocity relative to the planet, while those in the remaining panels trace the mean in-plane magnetic field.
}
\label{fig:mid_midplane}
\end{figure*}

\begin{figure*}[t!]
\centering
\includegraphics[trim=0mm 0mm 0mm 0mm, clip, width=7.in]{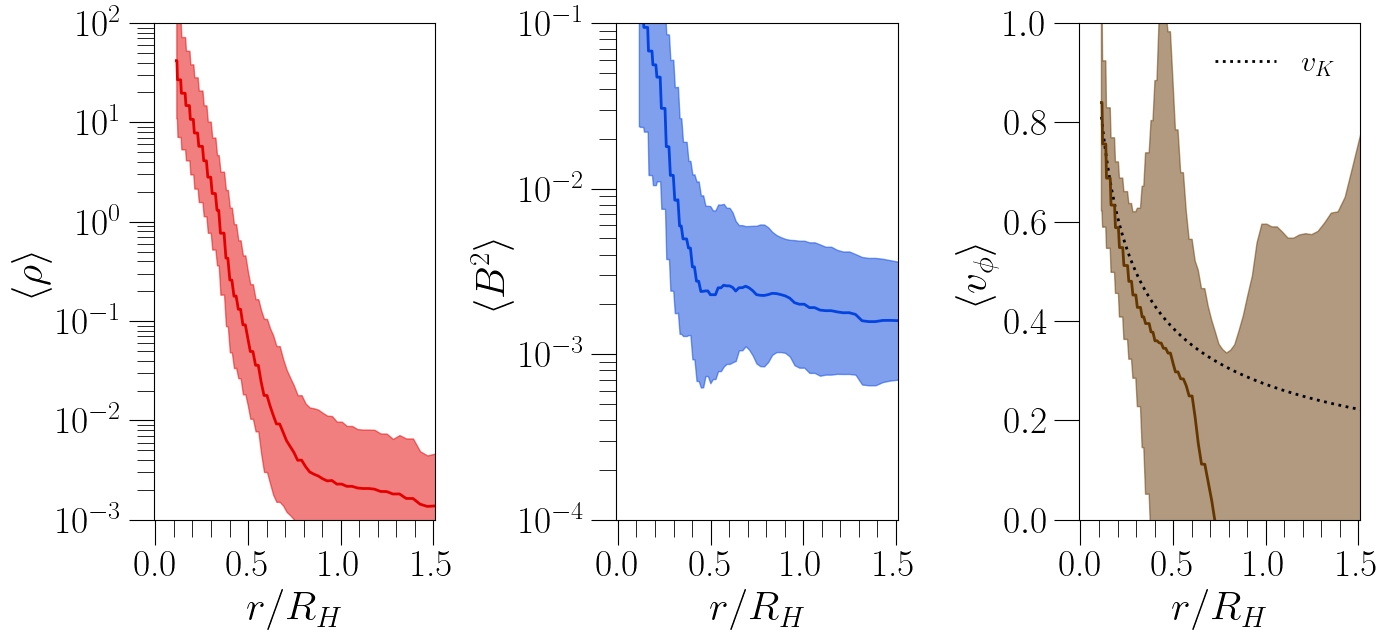}
\caption{
One-dimensional radial cuts for the \textit{middle planet} in the planet-centered frame, shown as functions of distance from the planet in units of the Hill radius, $R_H$. All profiles are azimuthally averaged at each snapshot and then time averaged the final \(2.5\,T_0\) interval.
Due to the polar orientation of the CPD, the azimuthal average is performed around the $+\hat{y}$ direction.
Solid lines show the mean profiles and shaded regions indicate the temporal scatter. From top left to bottom right, the panels show: the gas density $\rho$; the magnetic energy density $B^{2}$; and the azimuthal flow velocity $v_{\phi}$ together with the local Keplerian velocity $v_{K}$ (dotted line).
}
\label{fig:mid_linear}
\end{figure*}

We run the simulation for a total time of five orbits at the truncation radius (\(t = 5\,T_0\)) and obtain a total of 101 snapshots separated by \(\Delta t = 0.05\) intervals. 

Figure \ref{fig:three_planets} illustrates the midplane density structure and velocity at different times (left to right columns), focusing on different regions of the disk around each planet in detail. The top row shows a global view of the disk in the star's reference frame, with the positions of the planets highlighted.
Each colored square, centered on the corresponding planet, is drawn with a side length equivalent to four times the respective planet's Hill radius. From farthest to closest, these Hill radii are 0.269, 0.134, and 0.067 $R_0$, respectively.
The subsequent rows consist of zooms-ins into the local environment surrounding each planet with gas velocity plotted in their respective planet-centric corotating reference frames. These showcase the evolution of circumplanetary structures, with the velocity streamlines providing insight into the CPD flow patterns.

Starting from the second row in Figure \ref{fig:three_planets}), the \textit{outer planet} shows a CPD: The ordered counter-clockwise flow and high-density concentrations suggest it as stable, rotating, coplanar, and symmetric.
Next, in the third row, the \textit{middle planet}, also shows a CPD, but oriented perpendicular to the PPD.
Finally, in the last row, in contrast to its siblings, the \textit{inner planet} fails to form any lasting structure. Even though the velocity streamlines reveal the presence of the planet, by the next snapshot the planet's surroundings are disrupted and no permanent structure can develop.
In the following subsections we will look into the properties of these structures more closely.

\subsection{Angular-momentum Vector}

To quantify the net rotation of the gas around each planet, we follow the planet along its orbit and transform the surrounding flow into an instantaneous planet-centered comoving frame.

We define the planet-centered position and comoving velocity as
\begin{equation}
\mathbf{r}' \equiv \mathbf{r}-\mathbf{r}_p(t),
\label{eq:rprime_def}
\end{equation}
and
\begin{equation}
\mathbf{v}_{\rm com} \equiv \mathbf{v}-\mathbf{v}_p(t),
\label{eq:v_com}
\end{equation}
respectively.

The calculation is performed within a planet-centered Cartesian box extending $\pm0.6\,R_H$ along each coordinate axis (corresponding to a cube of side length $1.2\,R_H$).
Cells located within an inner spherical region of radius $0.1\,R_H$ are excluded in order to reduce the influence of the softened planetary potential and the rapidly varying flow near the planet \citep[see, e.g.,][]{Ben_tez_Llambay_2015}.

The total gas angular-momentum vector is computed as
\begin{equation}
\mathbf{L}(t) \;=\; \sum_{\rm cells} \left(\rho\,\Delta V\right) \left(\mathbf{r}'\times\mathbf{v}_{\rm com}\right),
\label{eq:L_planet}
\end{equation}
where $\rho$ and $\Delta V$ are the density and volume of each computational cell.

Figure \ref{fig:ang_mom} shows the time evolution of the Cartesian components of $\mathbf{L}$ for the three planets.
For the \textit{outer planet}, the angular momentum is strongly dominated by the vertical component, $L_z$.
After the planet is introduced, $L_z$ grows rapidly, reaches a maximum near $t\simeq67$--$68\,T_0$, and later decreases while remaining the dominant component throughout the simulation.
This decline is closely related to a  decrease in the amount of gas mass within the measurement region, consistent with fluctuations in availability associated to low-density, magnetic ``bubbles'' discussed in Paper I.
The persistent, comparative weakness of components $L_x$ and $L_y$ indicates that the rotation axis of the circumplanetary gas remains closely aligned with that of the parent PPD during the entire simulation interval, consistent with the presence of a coplanar CPD.
This is further visualized in in the top row of Figure \ref{fig:ang_mom_3D}.

The \textit{middle planet} exhibits a markedly different behavior with its total angular momentum evolution becoming dominated by the in-plane component $L_y$, which steadily grows over the simulated interval.
This implies that the net rotation axis of the circumplanetary gas lies approximately parallel to the PPD midplane, consistent with a polar CPD configuration.
This behavior is first seen in the third row of Figure \ref{fig:three_planets} and visualized more clearly in the bottom row of Figure \ref{fig:ang_mom_3D}, where the orientation of the circumplanetary structure remains preferentially oriented toward $+\hat{y}$ over the simulated interval as the planet orbits the star.
The remaining components, $L_x$ and $L_z$, have non-negligible magnitudes and fluctuate in time.

In contrast, the \textit{inner planet} maintains an almost zero and highly variable angular momentum signal, indicating the absence of a stable, rotationally supported CPD.
Instead, gas near the planet is confined to transient, envelope-like structures that repeatedly form and disperse.

\subsection{Coplanar CPD --- Outer Planet}

A long-lasting coplanar CPD forms around the \textit{outer planet}.
Its persistence is supported by the simulation in Paper I, in which a planet of the same mass in the same disk sustained a coplanar CPD over 30 planetary orbits.
Figure \ref{fig:slice_p2} shows an instantaneous \(x\)--\(z\) slice of the structure both within the PPD and in the immediate vicinity of the planet. In the zoomed view, the CPD appears as a thin, dense equatorial enhancement. The negative \(v_z\) above the midplane and positive \(v_z\) below it indicate gas converging vertically toward the CPD.

To average the CPD properties over time, we switch from the comoving frame defined above to a planet-centered, orbit-corotating frame. For each snapshot, we construct planet-centered maps in spherical-polar coordinates using the shifted position $\mathbf{r}'$ defined in Equation \ref{eq:rprime_def}.

To preserve long-lived asymmetries relative to the star, we rotate each snapshot so that the star--planet axis remains fixed, with the star always located west of the planet (\(\phi=180^{\circ}\)).
The velocity in this corotating frame is obtained from the comoving velocity by subtracting the rigid rotation associated with the planet's orbital motion: \begin{equation}
\mathbf{v}_{\rm cor}(\mathbf{r}',t) =
\mathbf{v}_{\rm com}(\mathbf{r}',t) -\mathbf{\Omega}_p(t)\times\mathbf{r}', \label{eq:v_corot}
\end{equation}
where $\mathbf{\Omega}_p(t)$ is the planet's instantaneous orbital angular velocity.

To construct the planet-centered midplane maps (Figure \ref{fig:out_midplane}), we take the slice at \(\theta=\pi/2\) from each spherical snapshot and time-average over the final $2.5\,T_0$ of the analyzed interval.
For the poloidal maps (Figure \ref{fig:out_poloidal}), we first azimuthally average and then time-average over the same interval.
We also construct one-dimensional radial profiles (Figure \ref{fig:out_linear}) from azimuthally and time-averaged cuts through the planet-centered midplane.

From these maps and profiles we observe that the \textit{outer planet} hosts a dense, long-lived, rotationally supported coplanar CPD.
The clearest dynamical evidence comes from the \(v_\phi\) panel of Figure \ref{fig:out_linear}, which shows that inside \(\approx 0.25\,R_H\) the azimuthal velocity follows the local planet-centered Keplerian profile before becoming increasingly sub-Keplerian farther out.
The corresponding spatial structure is shown in Figure \ref{fig:out_midplane}, where the midplane density map shows a compact enhancement centered on the planet. 
This dense region is threaded by closed, nearly circular streamlines, indicating that gas circularizes into a disk-like structure.
From the extent of the closed streamlines and the dense central enhancement, the main body of the CPD occupies the inner \(\approx 0.3\)–\(0.4\,R_H\), which is roughly consistent with the measured CPD sizes for simulated, super-thermal gas giants \citep[see, e.g.,][]{Martin_2011,Fung_2019,Li_2023}.
Outside this region, the circumplanetary flow reconnects to the ambient PPD through asymmetric spiral features, including a broader and more persistent starward arm and a weaker anti-starward arm that is more strongly smeared by time-averaging.
In the azimuthally averaged poloidal view of Figure \ref{fig:out_poloidal}, the CPD appears as a dense equatorial lobe centered on the midplane, extending radially outward and flaring into a broader skirt-like structure up to \(\approx 35^\circ\) from the horizontal.
Although azimuthal averaging washes out the spiral features visible in the midplane, it preserves the strong contrast between the dense low-latitude CPD and the diffuse high-latitude circumplanetary atmosphere.

The CPD is fed from high latitudes and participates in a large-scale meridional recycling flow.
The mean poloidal velocity field, \(\langle v_p\rangle/v_K\), shows gas entering the Hill sphere primarily from latitudes more than \(\sim 30^\circ\) from the midplane and leaving through the equatorial region.
The poloidal streamlines curve inward from the polar atmosphere toward the planet and are then redirected outward through the midplane, with part of that gas bending back toward higher latitudes farther out.
Consistent with this picture, the radial-velocity map shows inward motion at high latitudes (\(\langle v_r\rangle/v_K<0\)) close to the free-fall speed and outward motion in the dense equatorial lobe (\(\langle v_r\rangle/v_K>0\)).

The mass transport is vertically stratified, with decretion through the CPD midplane and accretion through the atmosphere.
As shown in the mass-flux map, \(2\pi r^2\sin\theta\,\langle \rho v_r\rangle\), the strongest positive flux is confined to a narrow equatorial layer spanning approximately \(\pm 15^\circ\) about the midplane.
By contrast, the negative mass flux associated with inflow is broader and more diffuse, only becoming concentrated into conical channels on final approach to the planet at roughly \(\pm 35^\circ\) from the midplane.

Although rotational support defines the inner CPD, meridional motions remain dynamically important throughout the surrounding circumplanetary flow.
The map of \(\langle v_p\rangle/v_K\) shows that the strongest poloidal motions occur close to the planet and along the polar inflow channels, where the poloidal speed reaches a substantial fraction of the local Keplerian speed.
At the same time, the velocity-angle map,
\(\arctan(\langle v_\phi\rangle/|\langle v_p\rangle|)\),
shows that the dense circumplanetary region retains a substantial prograde azimuthal component.
Over the CPD body, the angle remains moderately positive, indicating that both rotation and meridional circulation contribute to the mean flow, while toward the surface and polar regions the flow becomes increasingly poloidal.

This flow structure is in general agreement with previous, local and global, 3-D hydrodynamic simulations of CPDs \citep[see, e.g.,][]{Machida_2008,Tanigawa_2012,Fung_2019, Li_2023, Gardner_2026}.

The magnetic field is strongly amplified within the CPD and wound into a predominantly azimuthal configuration, as expected in magnetized flows subject to differential rotation \citep[e.g.,][]{Gressel_2013}.
In the midplane and poloidal maps, as well as in the radial profiles, the magnetic energy (\(B^2=B_r^2+B_\phi^2+B_\theta^2\)) is enhanced by several orders of magnitude in the same equatorial lobe that marks the CPD.
This enhancement is not perfectly axisymmetric, with the south-east side being somewhat stronger than the rest.
Figure \ref{fig:out_linear} further shows that within \(\sim 1\,R_H\), \(B_\phi\) exceeds \(B_\theta\), indicating that the field is being sheared and wound by circumplanetary rotation and by the surrounding recycling flow.

Despite the strong magnetic amplification, the CPD itself remains mostly gas-pressure dominated, while the surrounding ambient flow is more strongly magnetized.
The gas-to-magnetic pressure ratio is defined as $\beta \equiv 2P/B^2$.
Maps of this quantity show that most of the dense CPD has \(\beta \gtrsim 1\), even though the magnetic energy is strongly enhanced there.
In the midplane view, the \(\beta=1\) contour is distorted rather than circular and bulges toward the starward arm, reflecting the asymmetric magnetic amplification.
In the poloidal view (Figure \ref{fig:out_poloidal}), this contour encloses the dense CPD body, including the inflow channels, and extends somewhat beyond the Hill radius because the azimuthal average captures the spiral-arm contribution.
In the radial profiles, \(\beta_\theta\) remains much larger than the total \(\beta\), indicating that the vertical field component alone does not dominate the magnetic pressure budget within the CPD body.

The magnetic geometry changes from a toroidally wound field inside the CPD to a more weakly toroidal field in the surrounding atmosphere.
The poloidal magnetic-field angle map shows a moderately positive angle in the dense equatorial lobe, indicating a coherent azimuthal field component within the CPD.
Away from the CPD, the angle becomes weakly negative over most of the domain, implying a reversal of the mean toroidal field and a more nearly poloidal geometry in the ambient flow.

Magnetic torques are strongest in the radial direction within the CPD.
The Maxwell stress component \(-\langle B_rB_\phi\rangle\) is positive and concentrated in the equatorial lobe, with its largest values close to the planet and along both the dense midplane outflow and the \(\pm 30^\circ\) inflow cones.
This indicates that radial magnetic torques contribute to removing angular momentum from the CPD, similar to what occurs in MHD circumstellar disk \citep[see, e.g.][]{Zhu_2018}.
By contrast, \(-\langle B_\theta B_\phi\rangle\) divides into two radial regimes: between \(r \approx 0.3\) and \(0.6\,R_H\), angular momentum is transported from the atmosphere toward the midplane, whereas for \(r \lesssim 0.3\,R_H\) the direction reverses.
These torques will be further discussed in \S\ref{sec:disc}.

The integrated stress parameter \(\alpha_{\mathrm{int}}\) in Figure \ref{fig:out_linear} marks a dynamical transition near \(r\sim 0.3\,R_H\).
Starting from a high value in the inner CPD, the profile decreases to a local minimum at roughly the same radius where \(v_\phi\) departs from the planet-centered Keplerian profile.
Farther out, \(\alpha_{\mathrm{int}}\) rises again to values of order unity, but with substantially larger temporal scatter.
From that we interpret the inner \(r \lesssim 0.3\,R_H\) region as a comparatively coherent CPD with efficient angular-momentum transport, whereas beyond that radius the measured stresses increasingly reflect the lower-density, turbulent ambient flow of the parent PPD. Nevertheless, the CPD is efficient at transporting angular momentum with $\alpha_{\mathrm{int}}\gtrsim 0.5$.

\subsection{Polar CPD --- Middle Planet}

A circumplanetary structure also develops around the \textit{middle planet}, but its geometry is fundamentally different from the coplanar case.
Additionally, it also contrasts with the coplanar CPD found in the simulation of Paper I around a planet of the same mass at a similar location (\(r=1\,R_T\) instead of \(0.9\,R_T\)).
Figure \ref{fig:slice_p1} shows the instantaneous \(x\)--\(z\) slice of the structure both within the PPD and in the immediate vicinity of the planet. In this case, the slice cuts approximately through the plane of the polar CPD as we look at it face-on. The reversal in the sign of \(v_z\) across the planet is consistent with rotation about the \(y\)-axis.

As before, we construct planet-centered maps by shifting the origin to the instantaneous planet position. Unlike the coplanar case, however, we do not rotate snapshots to fix the star--planet axis prior to averaging, meaning we remain in comoving frame. This choice is motivated by the approximately persistent alignment of the circumplanetary angular-momentum vector with $+\hat{y}$ (see third row of Figure \ref{fig:three_planets} and second row of Figure \ref{fig:ang_mom}), so averaging in the nonrotating, planet-centered frame preserves the orientation of the polar CPD.
We examine two perpendicular cuts: the orbital-midplane cut, which shows the polar CPD approximately edge-on, and a cut normal to $+\hat{y}$, which reveals it approximately face-on. The latter is constructed by averaging the two adjacent azimuthal cell layers on either side of $Y=0$, using the same fixed orientation throughout the averaging interval.

The time-averaged maps in Figure \ref{fig:mid_midplane} show complementary views of the polar CPD near the disk--cavity interface, with the circumplanetary material lying above and below the disintegrating PPD midplane.
In the upper row, the orbital-midplane cut reveals an edge-on, two-lobed density enhancement extending to approximately $0.3\,R_H$ on opposite sides of the planet, with comparatively depleted regions along the projected polar direction.
As with its coplanar sibling, the velocity streamlines show gas entering the planet's vicinity primarily from its poles rather than through its equator.
However, meridional circulation is not as directly discernible.
In the lower row, the perpendicular cut reveals a centrally concentrated, nearly circular structure with velocity streamlines circulating around the planet.

The magnetic field is also strongly amplified around the polar CPD. Both view make this central enhancement clearer, while the magnetic streamlines exhibit a complex geometry.
As in the coplanar case, the dense inner CPD remains gas-pressure dominated (\(\beta>1\)) despite its location within a magnetically dominated cavity.
Comparing the radial profiles in Figure \ref{fig:mid_linear} with those from Figure \ref{fig:out_linear} shows that, for both planets, the radial density and magnetic energy behave very similarly, having virtually the same magnitude close to the planet and then sharply dropping, reaching close to a background level by \(r \sim 0.5\,R_H\).
Because of the magnetospheric-accretion environment around the \textit{middle planet}, the background \(\langle\rho\rangle\) is at least one order of magnitude lower, while \(\langle B^2\rangle\) is at least one order of magnitude higher, than around the \textit{outer planet}.
At the same time, these quantities show larger temporal variability for the \textit{middle planet}, reflecting the strongly time-dependent cavity flow surrounding it \citep[see, e.g.,][]{Zhu_2024,Zhu_2025}.

Consistent with the circulation visible in the face-on maps, the azimuthal velocity profile shows that the polar CPD possesses a rotating inner structure, although one that is less regular than in the coplanar case.
Near the planet, \(r \lesssim 0.2\,R_H\), the circumplanetary gas reaches azimuthal speeds close to the planet-centered Keplerian around the planet.
Farther out, however, the profile becomes more irregular and more strongly affected by temporal variability, and by \(r \sim 0.6\,R_H\) any semblance of circulation breaks down, reflecting the transition from the bound structure to the surrounding circumstellar disk flow.
Thus, compared with the \textit{outer planet}, the \textit{middle planet} hosts a circumplanetary disk that is more weakly isolated from its environment, more strongly influenced by the anisotropic magnetic and vertical inflow, and yet still able to preserve a persistent polar orientation.

\section{Discussion}\label{sec:disc}

\begin{figure}[t!]
\centering
\includegraphics[trim=0mm 0mm 0mm 0mm, clip, width=\columnwidth]{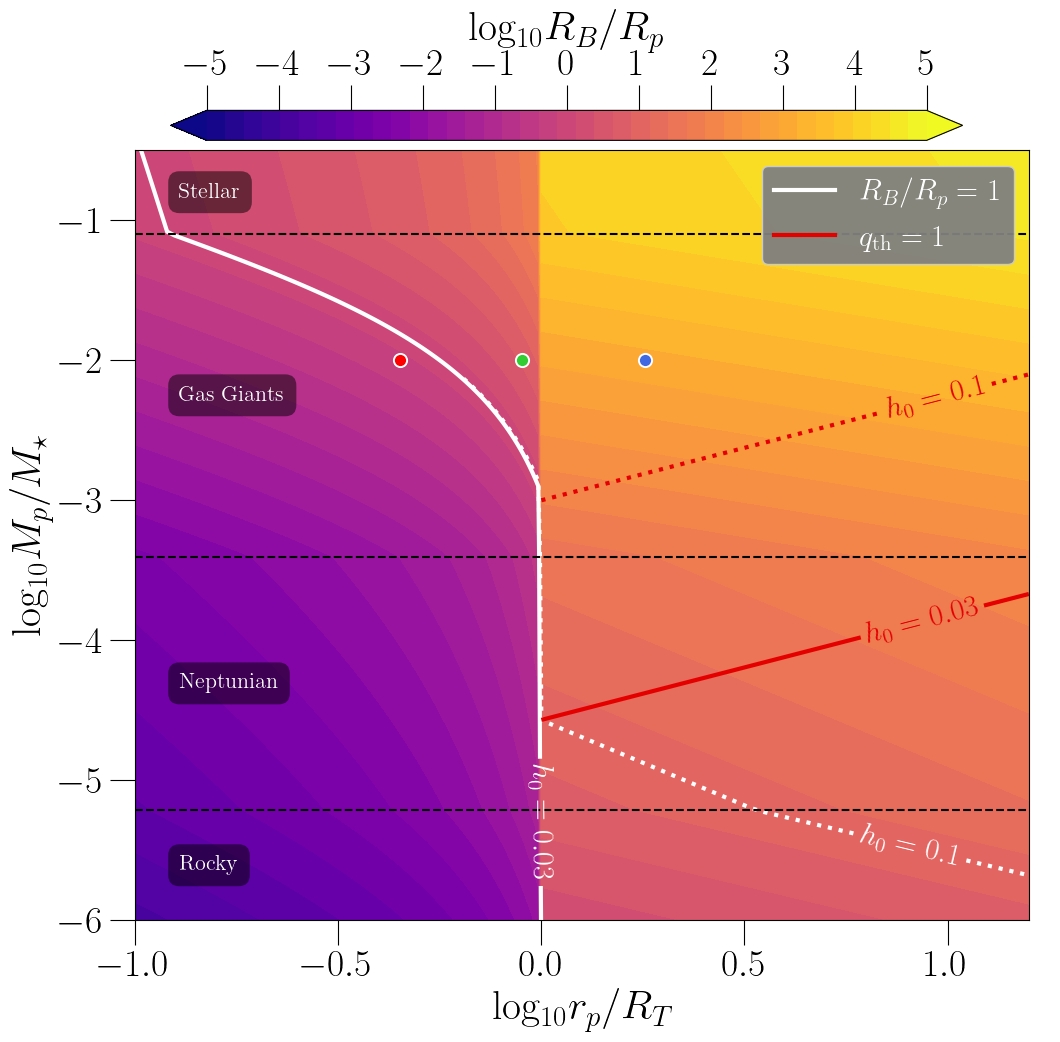}
\caption{
Heatmap of the ratio of Bondi radius to planet radius, \(R_B/R_p\), as a function of planet mass and distance from the star.
The magnetospheric truncation radius, \(r_p=R_T=10\,R_\odot=0.047\,{\rm au}\), separates the magnetosphere (\(r_p<R_T\)) from the turbulent inner disk (\(r_p>R_T\)), where we adopt the thin-disk aspect ratio profile \(h = 0.03 (r/R_T)^{1/4}\).
Inside the magnetosphere, the effective Bondi radius from Equation \ref{eq:bondi_eff} is used.
The horizontal dashed lines mark broad planetary mass--radius regimes, based on Equation \ref{eq:mass_radius_relation} and Table \ref{tab:mass_radius_k} of \citet{Chen_2016}.
The white curves delineate \(R_B/R_p=1\), where the solid one corresponds to the thin-disk case (\(h_0=0.03\)) and the dotted one to the thick-disk case (\(h_0=0.1\)).
In the disk region, the red curves indicate the \(q_{\rm th}=1\) boundary, with the solid and dotted lines corresponding to the same thin and thick cases, respectively.
The red, green, and blue dots mark the simulated \(q=0.01\) planets, corresponding to \(M_p\simeq10\,M_J\), at \(r_p=\{0.45,0.90,1.80\}R_T=\{0.021,0.042,0.084\}\,{\rm au}\).
}
\label{fig:bondi}
\end{figure}

Similar to Paper 1, we set the code length unit \(R_0\) equal to the magnetospheric truncation radius,
\(R_T=10\,R_\odot=0.047\,{\rm au}\), around a Solar analogue with \(M_\star=1\,M_\odot\). The corresponding time unit is \(1/\Omega_0=1.6\times10^{-3}\,{\rm yr}\). Equating the mean code accretion rate of the disk, \(\dot{M}=-0.005\), with a physical accretion rate of \(\dot{M}=10^{-8}\,M_\odot\,{\rm yr}^{-1}\) gives a mass unit of \(3.19\times10^{-9}\,M_\odot\), and therefore a surface-density unit of \(13.1\,{\rm g\,cm^{-2}}\).

With this normalization, the three simulated planets with \(q=0.01\) correspond to \(M_p\simeq10\,M_J\). Their orbital radii \(r_p=\{0.45,0.90,1.80\}R_T\) correspond to \(\{0.021,0.042,0.084\}\,{\rm au}\). Their Hill radii are
\(R_H=r_p(q/3)^{1/3}=\{0.067,0.134,0.269\}R_T\), or \(\{0.0031,0.0063,0.0125\}\,{\rm au}\), respectively.

\subsection{CPD Formation Condition}

\begin{table*}[t!]
\centering
\caption{Mass-radius relation parameters by planetary regime. Adapted from \cite{Chen_2016}.}
\label{tab:mass_radius_k}
\begin{tabular}{lccc}
\hline
\textbf{Planet Regime} & \textbf{Mass Range} \((M_\oplus)\) & \textbf{Coefficient \( C \)} & \textbf{Exponent \( k \)} \\
\hline
Rocky & \( M_p < 2.04 \) & 1.01 & 0.28 \\
Neptunian & \( 2.04 < M_p < 132 \) & 0.81 & 0.59 \\
Gas Giant & \( 132 < M_p < 2.66 \times 10^4 \) & 17.8 & -0.04 \\
Stellar & \( M_p > 2.66 \times 10^4 \) & $1.41 \times 10^{-3}$ & 0.88\\
\hline
\end{tabular}
\end{table*}

In our simulation, the three planets have the same mass, but orbit at very different locations. Two of them develop CPDs, whereas the innermost one does not.
Although this inner planet appears to intermittently capture gas from accretion columns that cross the cavity, the captured material does not settle into a persistent envelope, let alone a long-lived CPD.
As a minimum requirement for accretion, the planet must have a Bondi radius or Bondi–Hoyle–Lyttleton radius that extends beyond its physical radius, \(R_p\), so that
\begin{equation}
    \frac{R_B}{R_p} > 1.
\label{eq:bondi_cond}
\end{equation}

Within the circumstellar disk, where the gas rotates close to the local Keplerian speed, the relative velocity between an orbiting planet and the average ambient flow is expected to remain modest, so that $v_{\rm rel}\ll c_s$ and the standard Bondi radius of Equation \ref{eq:bondi_basic} remains a reasonable estimate.
Inside the magnetospheric cavity, however, the gas is comoving with the stellar magnetosphere while being funneled and free-falling onto the star, acquiring a substantial velocity difference to the planet.

The Bondi–Hoyle–Lyttleton radius
\begin{equation}
R_{B,\mathrm{eff}} = \frac{G M_p}{c_s^2 + v_\mathrm{rel}^2}
\label{eq:bondi_eff}
\end{equation}
can be much smaller than the usual Bondi radius when $v_{\rm rel}\gtrsim c_s$.
In the limiting case of a non-rotating magnetosphere, as in our simulation, \(v_{\rm rel}\approx v_K \gg c_s\).
More generally, if the star and magnetosphere have a non-zero spin and rotate with angular frequency \(\Omega_\star\), we estimate
\begin{equation}
v_{\rm rel} \approx v_K - \Omega_\star r_p,
\label{eq:vrel_cavity}
\end{equation}
where \(\Omega_\star = \sqrt{GM_\star/R_C^3}\) and \(R_C\) is the stellar corotation radius.
Considering young stars that are in the spin-equilibrium state, 3-D magnetospheric accretion simulations yield \(R_T \simeq 0.79\,R_C\) \citep{Zhu_2025}.

To compare $R_B$ with the physical planet size, we estimate $R_p$ from the empirical mass--radius relation of \citet{Chen_2016},
\begin{equation}
    \frac{R_p}{R_{\oplus}} = C \left( \frac{M_p}{M_{\oplus}} \right)^k,
    \label{eq:mass_radius_relation}
\end{equation}
where \(R_{\oplus}\) and \(M_{\oplus}\) are Earth's radius and mass, and the empirical coefficients $(C,k)$ depend on the relevant mass regime. These are listed in Table \ref{tab:mass_radius_k}.

Combining Equations \ref{eq:bondi_cond}--\ref{eq:mass_radius_relation} with a more realistic thin inner-disk aspect ratio profile \(h = 0.03 (r/R_T)^{1/4}\), rather than the thicker disk used in our simulation, we obtain Figure \ref{fig:bondi}.
This is a heatmap of the ratio \(R_B/R_p\) across planet mass and radial distance.
The solid white contour marks $R_B/R_p=1$ inside the disk and the magnetospheric cavity, while the dotted one is for a thick-disk case with $h_0=0.1$.
The colored dots mark the masses and orbital locations of the simulated planets.
We observe that this condition is quite restrictive in the cavity, while in the disk it is easily satisfied throughout the whole parameter space in the thin-disk case and only restricts low-mass planets for the thick-disk.
In our setup, the \textit{inner planet} (red dot) lies below the \(R_B/R_p=1\) threshold, consistent with its failure to maintain a persistent circumplanetary structure.
By contrast, the \textit{middle planet} and especially the \textit{outer planet} satisfy this minimum capture condition.

A second useful diagnostic is the ratio of planet mass to the local thermal mass, \(q_{\rm th}\), defined in Equation \ref{eq:qth}.
Once in the super-thermal regime, \(q_{\rm th}>1\), the planet is massive enough to strongly perturb the surrounding gas favoring the development of a disk-like circumplanetary flow \citep[see, e.g.,][]{Fung_2019, Sagynbayeva_2025}.
This condition is still not sufficient by itself: sub-thermal planets may retain a compact inner Keplerian region, while even super-thermal planets can remain envelope-dominated if cooling is inefficient \citep[see, e.g.,][]{Szulagyi_2017, Schulik_2020, Krapp_2024}.
In our simulation, the near-instantaneous thermal relaxation should make bound gas more likely to settle into a CPD than in adiabatic or slowly cooling cases.

In Figure \ref{fig:bondi}, the red curves mark the \(q_{\rm th}=1\) threshold for the two representative aspect ratios.
For the thick-disk case ($h_0=0.1$), the super-thermal regime is limited to massive gas giants, whereas for the thin-disk case ($h_0=0.03$) it extends to much lower masses.
In this framework, our \textit{outer planet} satisfies both the Bondi condition and the thermal-mass condition, with \(q_{\rm th} \approx 6.43\), consistent with the rapid development of a stable, rotationally supported, coplanar CPD.

The \textit{middle planet} occupies a more ambiguous regime.
If treated as fully embedded in the cavity, then the disk-based interpretation of \(q_{\rm th}\) becomes less straightforward.
However, for non-rotating and slowly rotating stars, the magnetically supported cavity is subject to ``interchange instability'', allowing for the magnetized gas filaments (``fingers'') to penetrate into it (see top row of Figure \ref{fig:three_planets}).
This process maintains an appreciable density in the midplane, dropping as the filaments are lifted up during their advance towards the star \citep[][]{Kulkarni_2008, Zhu_2024, Zhu_2025}.
In that sense, the feeding environment of the \textit{middle planet} still resembles the inner PPD more than an evacuated cavity, and the planet remains effectively in the super-thermal regime (with \(q_{\rm th}\sim 10\)).
This provides a natural explanation for why the planet can still sustain a circumplanetary disk despite orbiting just inside the truncation radius.

The polar orientation of that CPD is likely set by the angular momentum of the gas that is actually captured, rather than by the midplane orientation of the ambient PPD alone, while magnetic stresses may help the structure remain coherent once established (see further discussion in \S \ref{subsec:angmom}).
Near the truncation radius, the inflow is intrinsically anisotropic and shaped by magnetospheric accretion, so the gas entering the Hill sphere need not carry angular momentum aligned with the PPD rotation axis \citep[see, e.g.,][]{Zhu_2024, Zhu_2025}.
Stochastic accretion has previously been suggested as a source of circumplanetary misalignment \citep{Gressel_2013}.
\citet{Chametla_2026b} likewise discuss misaligned angular-momentum delivery and direct magnetic torques as possible contributors to the strong CPD tilt in their simulations.
Therefore, the preferential alignment of our CPD's angular-momentum vector with $+\hat{y}$ may reflect the particular feeding history and local magnetic geometry, rather than a generally preferred direction.

If sustained for long enough, a misaligned CPD may influence the spin evolution and obliquity of the forming planet by delivering angular momentum that is not aligned with the planetary orbit \citep{Martin_2020, Martin_2021}.
Additionally, the presence and geometry of a CPD influences how gas is delivered to the planet, which in turn affects the variability and observability of planetary accretion signatures \citep[e.g.,][]{Aoyama_2018, Brittain_2020}.

Whether this polar configuration is a long-lived equilibrium, however, is unclear.
Secular tidal torques from the star can drive tilt evolution in misaligned CPDs, and after the PPD dissipates, these detached disks may, in principle, enter Kozai--Lidov-like inclination--eccentricity cycles once their tilt becomes sufficiently large \citep[see, e.g.,][]{Lubow_2017,Martin_2014, Martin_2020, Martin_2021}.
Our system differs from those idealized cases because the CPD is not cleanly detached. Continuous feeding and the strongly magnetized environment could damp, reorient, or repeatedly rebuild the disk.
Determining whether the polar state is transient, recurrent, or genuinely preferred therefore requires longer-duration simulations that vary the planet location, thermodynamics, and magnetic-field configuration, along with dedicated analysis of inflow geometry and angular-momentum budgets.
We leave this to future work.

\subsection{Angular-Momentum Redistribution in the Outer CPD}\label{subsec:angmom}

\begin{figure*}[t!]
\centering
\includegraphics[trim=0mm 0mm 0mm 0mm, clip, width=6.0in]{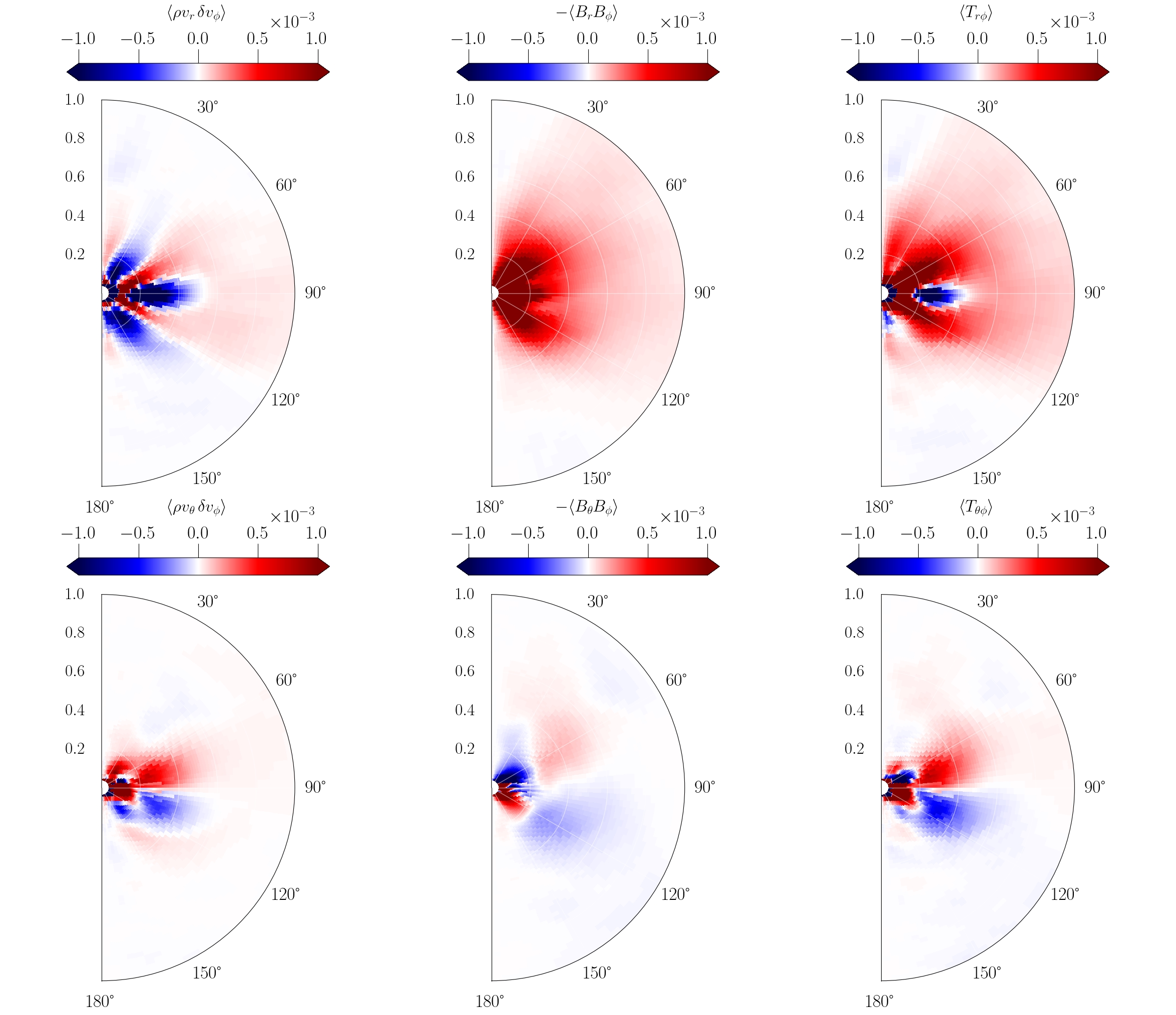}
\caption{
Poloidal maps of the azimuthally and time-averaged stresses \(\langle T_{r\phi} \rangle\) and \(\langle T_{\theta\phi}\rangle\) and their components from Equation \ref{eq:T_sph} in the planet-centered frame.
\textbf{Top row (left to right):} momentum (Reynolds) contribution $\langle \rho v_r\,\delta v_\phi\rangle$, magnetic (Maxwell) contribution $-\langle B_r B_\phi\rangle$, and total stress $\langle T_{r\phi}\rangle$.
\textbf{Bottom row (left to right):} momentum contribution $\langle \rho v_\theta\,\delta v_\phi\rangle$, magnetic contribution $-\langle B_\theta B_\phi\rangle$, and total stress $\langle T_{\theta\phi}\rangle$.
}
\label{fig:T-terms}
\end{figure*}

\begin{figure*}[t!]
\centering
\includegraphics[trim=0mm 0mm 0mm 0mm, clip, width=7.in]{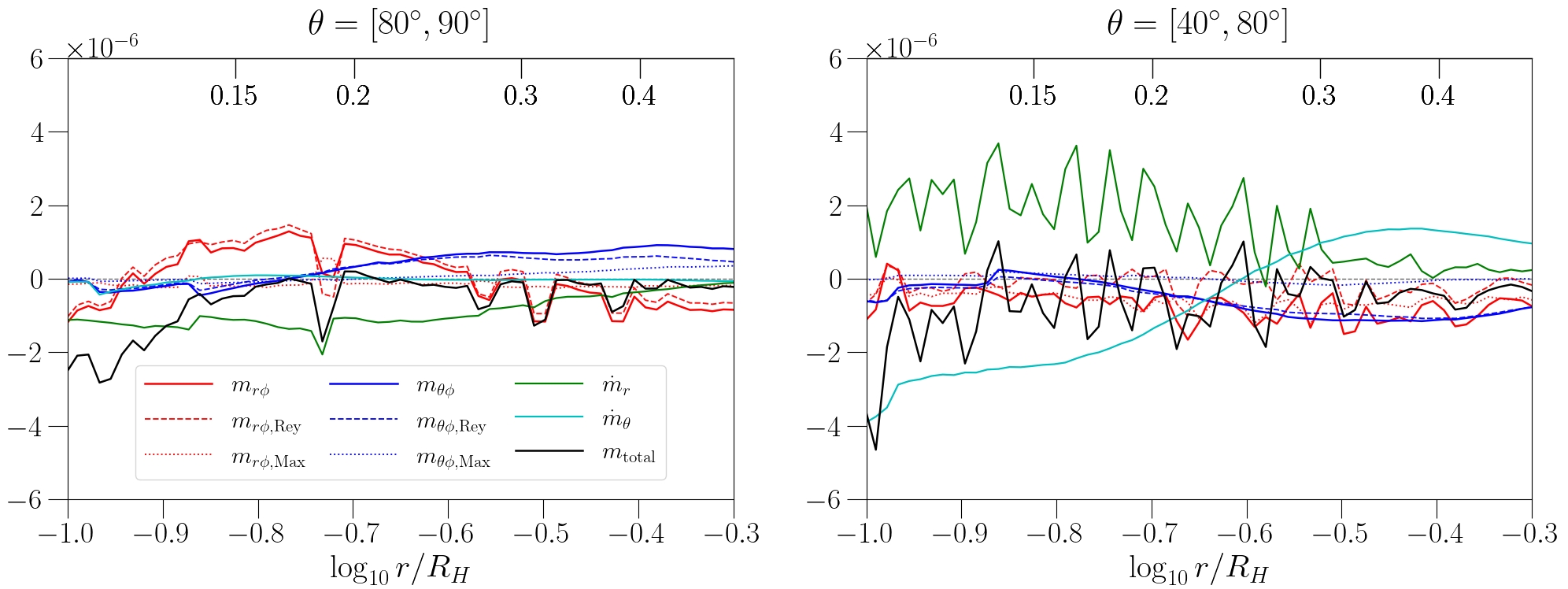}
\caption{
Angular-momentum budgets from Equation \ref{eq:m-terms} within different \(\theta\)-wedges. The time-averaged budget components are integrated over space (\(2\pi\) in \(\phi\), and \(\theta_{\mathrm{min}}\) to \(\theta_{\mathrm{max}}\) with the weight of the \(\sin^2\theta\) factor). The resulting quantities have been then multiplied by \(1/2\pi \times r^{3.5}\) so that they appear flattened along the radius-axis and have similar scales. The stress-gradient terms \(m_{r\phi}\) and \(m_{\theta\phi}\) are plotted both as totals and as separate contributions obtained by computing them with only the Reynolds or only the Maxwell part of the corresponding stress.
The bottom x-axis is written on logarithmic scale, while the top x-axis shows the corresponding linear radius \(r/R_H\).
}
\label{fig:m-terms}
\end{figure*}

We now study the mechanisms that lead to accretion and meridional circulation for the \textit{outer planet}'s CPD.
To do so, we evaluate the azimuthally and time-averaged stresses appearing in Equation \ref{eq:ang_mom_stress} in the same planet-centered frame used for Figure \ref{fig:out_poloidal}.
Figure \ref{fig:T-terms} shows the radial and meridional stresses, \(\langle T_{r\phi}\rangle\) and \(\langle T_{\theta\phi}\rangle\), together with their Reynolds and Maxwell contributions.

Radial and meridional stresses play different dynamical roles.
The radial stress is positive over most of the CPD and surrounding atmosphere, and its Maxwell component is much more extended than its Reynolds component.
This indicates that angular momentum is transported radially mainly by magnetic torques, not by the hydrodynamic correlations alone.
Only near the midplane around \(r\sim 0.2\)–\(0.3\,R_H\) does \(\langle T_{r\phi}\rangle\) show a local sign reversal, close to the transition where the inner CPD ceases to follow the planet-centered Keplerian profile.
The meridional stress map generally shows \(\langle T_{\theta\phi} \rangle > 0\) and \(\langle T_{\theta\phi} \rangle < 0\) on the upper and lower hemispheres, respectively, indicating the transport of angular-momentum from off-midplane layers into the midplane, and is dominated primarily by the Reynolds component.
In other words, the meridional circulation is mainly driven by hydrodynamical processes similar to previous hydrodynamical simulations \citep[e.g.,][]{Tanigawa_2012, Fung_2019, Li_2023}, while disk accretion is driven by magnetic fields.

With these stresses we then compute the angular-momentum budget terms in Equation \ref{eq:m-terms} and integrate them over two representative polar wedges spanning \(0.1\le r/R_H \le 0.5\) (Figure \ref{fig:m-terms}).
The near-midplane wedge, \(\theta=[80^\circ,90^\circ]\), corresponds to the dense CPD body.

There, the radial advection term \(\dot m_r\) is negative (radial outflow) and is offset by a positive \(m_{r\phi}\) over much of the inner disk and by a positive \(m_{\theta\phi}\) farther out.
Both of these gradients, \(m_{r\phi}\) and \(m_{\theta\phi}\), are dominated there by their Reynolds contributions.
The atmosphere wedge, \(\theta=[40^\circ,80^\circ]\), behaves differently: \(\dot m_r\) is the dominant positive term (radial inflow), while \(m_{r\phi}\) is mostly negative and \(m_{\theta\phi}\) remains comparatively weaker.
It's important to note that here \(m_{r\phi}\) is primarily magnetically driven.
\(\dot{m}_\theta\) completes the picture by transitioning from strongly negative in the inner disk to strongly positive in the outer disk, consistent with the descending flow at smaller radii and the returning ascent farther out.
In both wedges the total budget stays close to zero because of strong cancellations among the individual terms, indicating that over the analyzed interval the CPD is in a quasi-steady recycling state rather than undergoing a secular build-up of angular momentum.

This physical configuration is qualitatively consistent with earlier CPD studies that emphasized vertical inflow and equatorial outflow \citep[e.g.,][]{Tanigawa_2012, Gressel_2013, Li_2023}.
It also shows that magnetic torques can contribute significantly to the angular-momentum budget of a magnetized CPD.
Although coherent jets or winds can in principle be launched from magnetized CPDs when the field geometry and mass loading are favorable \citep[e.g.,][]{Pudritz_2007, Gressel_2013}, we do not observe such an outflow here (see vertical-velocity panel in Figure \ref{fig:slice_p2}).
High-resolution ideal-MHD simulations by \citet{Chametla_2026a} found that CPD jets are launched close to the planet, primarily from within $0.02 \lesssim r/R_H \lesssim 0.08$.
Such jets can develop even in highly tilted CPDs \citep{Chametla_2026b}.
For the \textit{outer planet} and \textit{middle planet}, the adopted smoothing lengths, $r_s\simeq0.045\,R_H$ and $r_s\simeq0.089\,R_H$, respectively, overlap this interval.
In our current analysis, we treat the flow within $0.1,R_H$ as numerically uncertain.
We may therefore fail to capture the conditions required for jet launching, and the absence of a detected jet does not exclude its development in this environment.
Future simulations that resolve the planetary boundary and innermost CPD will be needed to address this possibility.

\subsection{Mass Inflow and Planet Growth Timescales}

\begin{figure}[t!]
\centering
\includegraphics[trim=0mm 0mm 0mm 0mm, clip, width=\columnwidth]{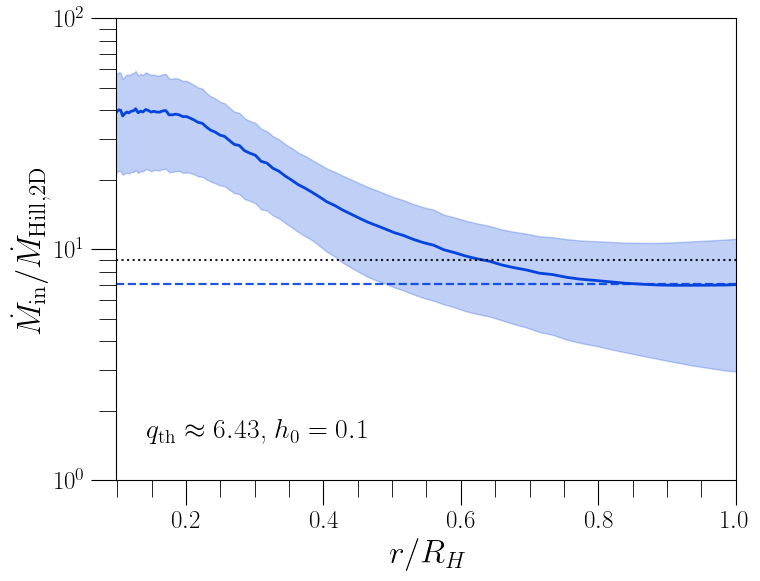}
\caption{
Time-averaged one-way mass inflow rate, $\dot{M}_{\rm in}$, through planet-centered spherical surfaces around the \textit{outer planet}.
The rate is normalized by the two-dimensional Hill accretion scale, $\dot{M}_{\rm Hill,2D}$. The shaded region indicates the temporal scatter over the averaging interval.
The dashed horizontal line marks the measured plateau level, computed over $0.8 \leq r/R_H \leq 1$.
The dotted horizontal line shows the sink-less two-dimensional reference value $\dot{M}_{p,\rm in}\simeq 9\dot{M}_{\rm Hill,2D}$ from \citet{Choksi_2023}.
}
\label{fig:mdot-in}
\end{figure}

\begin{figure}[t!]
\centering
\includegraphics[trim=0mm 0mm 0mm 0mm, clip, width=\columnwidth]{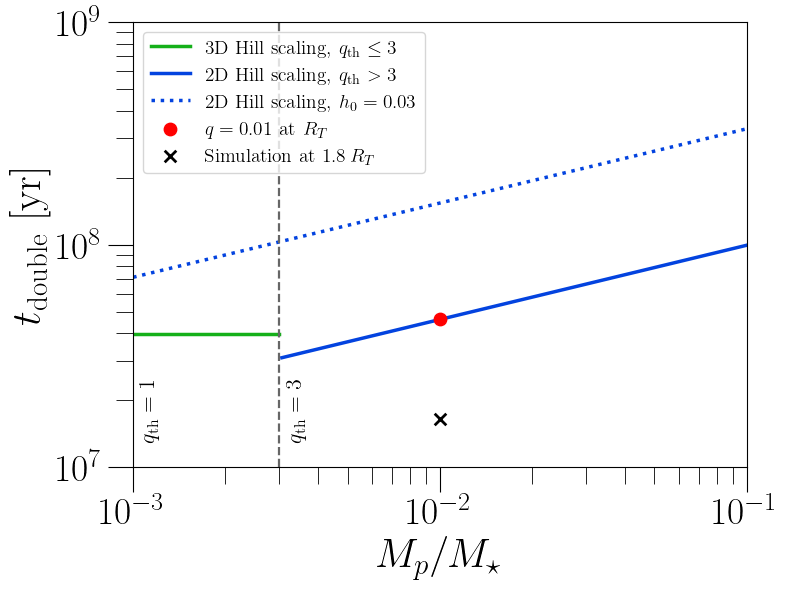}
\caption{
Estimated planet mass doubling timescale, $t_{\rm double}=M_p/\dot{M}_{p,\rm in}$, for planets located at the truncation radius, $R_T$, using the fiducial scaling \(\dot{M}=10^{-8}\,M_\odot\,{\rm yr}^{-1}\) and assuming the same ambient density and plateau inflow rate measured for the \textit{outer planet}.
For planets in the marginally super-thermal regime (\(1 \leq q_{\rm{th}} \leq 3\)), the inflow is normalized using the 3D Hill scaling, while planets in the more strongly super-thermal regime ($q_{\rm th} > 3$) use the 2D Hill scaling.
The red point marks the extrapolated value for a planet with $q=0.01$ placed at $R_T$, while the black cross shows the corresponding doubling time measured directly from the simulation at $1.8\,R_T$.
The vertical dashed line indicates the regime transitions at $q_{\rm th}=3$ for $h_0=0.1$.
}
\label{fig:double-time}
\end{figure}

\begin{figure*}[t!]
\centering
\includegraphics[trim=0mm 0mm 0mm 0mm, clip, width=6.in]{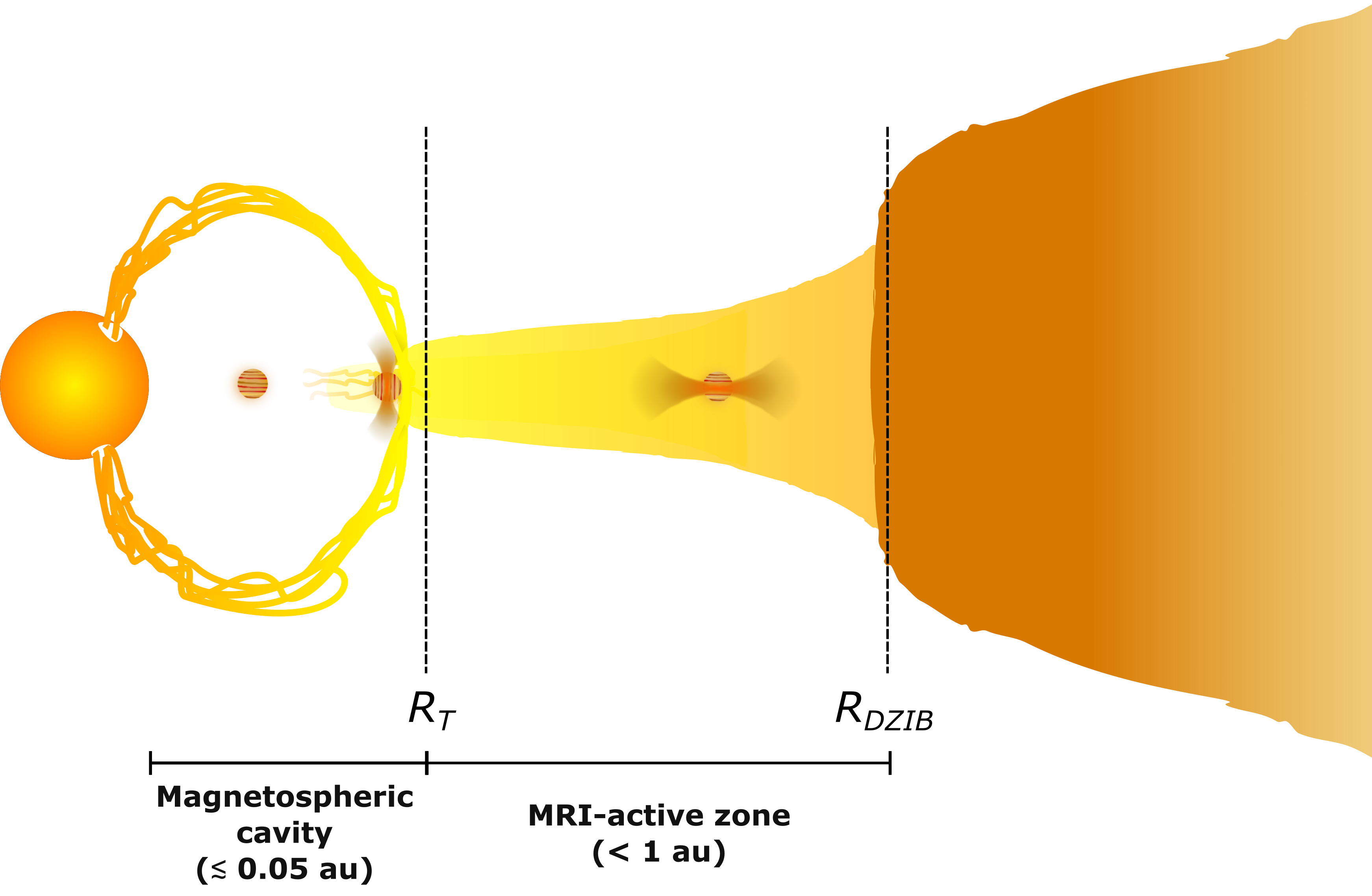}
\caption{
Schematic depiction of the PPD and the  circumplanetary environments of the three massive \(q=0.01\) planets at the end of the simulation. The \textit{outer planet}, embedded in the MRI-active inner disk, forms a rotationally supported CPD aligned with the PPD's midplane. The \textit{middle planet}, located near the truncation radius within the magnetospheric cavity, also forms a rotationally supported CPD, but with a polar orientation. The \textit{inner planet}, located closest to the star, fails to have any long-lived structure develop around it.
}
\label{fig:cpd_diagram}
\end{figure*}

To estimate the rate at which gas is supplied to the circumplanetary region and the maximum possible accretion rates, we follow the approach of \citet{Choksi_2023} and measure the one-way inflow rate through planet-centered spherical surfaces, $\dot{M}_{p,\rm in}$. 
This counts only gas with negative planet-centered radial velocity, $v_r<0$, and should therefore be interpreted as an upper limit to the true planetary accretion rate, since gas entering the Hill sphere may later leave unless it can cool and shed angular momentum.

In the marginally super-thermal regime (\(1 \leq q_{\rm{th}} \leq 3\)), where the Hill sphere remains embedded within the disk scale height, the inflow is approximately three-dimensional and can be normalized by
\begin{equation}
\dot{M}_{\rm Hill,3D} = R_H^3 \Omega_p \rho_0 ,
\end{equation}
where $\Omega_p$ is the planet's orbital frequency, and $\rho_0$ is the ambient midplane volumetric density near the planet.
In the strongly super-thermal regime (\(q_{\rm{th}} \gtrsim 10\)), the Hill sphere extends beyond the vertical thickness of the disk, so the effective cross-section becomes closer to $R_H H_p$ rather than $R_H^2$.
The corresponding two-dimensional Hill scaling is
\begin{equation}
\dot{M}_{\rm Hill,2D} = R_H^2 H_p \Omega_p \rho_0 .
\end{equation}

Our \textit{outer planet} has $q_{\rm th}\simeq 6.43$, placing it between the two regimes.
Despite the relatively thick disk in our simulation (\(h_0=0.1\)) and the planet's proximity to the truncation radius, Figure \ref{fig:mdot-in} shows a radial dependence and normalization closely resembling those reported by \citet{Choksi_2023}.
After decreasing toward small radii, the normalized inflow rate approaches a broad plateau by \(r/R_H \simeq 0.8\), within the \(0.4\lesssim r/R_H\lesssim 1\) interval reported by \citet{Choksi_2023}. Its normalization, \(\dot{M}_{p,\rm in}\simeq7\,\dot{M}_{\rm Hill,2D}\), is also close to their sink-less 2-D value of \(\simeq9\,\dot{M}_{\rm Hill,2D}\) (compare with their Figure 6).

That the inflow rate plateau differs from the \citet{Choksi_2023} scaling by only a factor of \(\approx 1.3\) suggests that Hill-based accretion estimates remain broadly applicable even for close-in planets embedded in magnetospherically influenced inner disks. This indicates that the gas supply rate into the Hill sphere is not significantly affected by the presence of magnetic fields. 
At the same time, the offset in both the plateau level and the radius at which it emerges indicates that the detailed inflow structure depends on the detailed geometry and temporal variability of the local disk/fields environment.
In particular, the presence of MHD turbulence and magnetic stresses may all modify how gas is funneled into the planet's Hill sphere compared to idealized hydrodynamic setups.

Using the measured plateau inflow rate for the \textit{outer planet} and the fiducial scaling, we estimate a characteristic mass doubling timescale of \(t_{\rm double}\approx1.65\times10^{7}\) yr for the simulated planet located at \(1.8\,R_T\).
To explore how this timescale varies with planet mass, Figure \ref{fig:double-time} extrapolates the inflow scalings to planets placed at the truncation radius, \(R_T\), assuming \(h_0=0.1\) and the same ambient density.
For planets in the marginally super-thermal regime, we adopt the 3-D Hill scaling, while more massive planets use the 2-D one.
We include the thinner-disk case (\(h_0=0.03\)) for comparison, which would remain strongly super-thermal over the plotted mass range. Since \(\dot{M}_{p,\rm in}\propto h_0\), the corresponding \(t_{\rm double}\) timescale increases by a factor of \(\simeq 3.3\).

Under these assumptions, super-thermal planets yield doubling timescales \(\gtrsim10^7\) yr.
Since protoplanetary disk lifetimes are typically estimated to range from a few to several tens of Myr \citep[see, e.g.,][]{Pfalzner_2024}, and because \(\dot{M}_{p,\rm in}\) is interpreted as the upper limit on the true mass accretion rate, \(\dot{M}_p\), these estimates suggest that giant planets residing in the magnetically active inner regions of PPDs may experience only limited late-stage growth.
Furthermore, a substantial fraction of the gas entering the Hill sphere may ultimately be recycled back into the surrounding disk rather than permanently accreted onto the planet.

\subsection{Model Caveats}

The simulations analyzed here share the same global setup as in Paper I, so many of the limitations discussed there also apply to the present work.
Although the simulations capture the global interaction between MRI-driven turbulence, magnetospheric accretion, and embedded giant planets, several assumptions should be kept in mind when interpreting the circumplanetary structures, dynamics and criteria presented here.

First, the stellar magnetic field is treated in an idealized manner.
We neglect stellar rotation and therefore do not self-consistently model the corotation radius or the angular-momentum exchange between the star and disk \citep[see][]{Zhu_2025}.
In addition, the stellar magnetic dipole is fixed along the simulation's \(z\)-axis and does not include oblique or multipolar components, despite observations suggesting that T Tauri stars often possess more complex magnetic topologies \citep[e.g.,][]{Donati_2007,Lankhaar_2022}.
Planetary and large-scale external magnetic fields are also ignored.
Such effects may alter, in a nontrivial manner, the geometry of the funnel flows, the structure of the inner disk, and the angular-momentum transport within the magnetosphere \citep[e.g.,][]{Romanova_2012,Romanova_2015,Parfrey_Tchekhovskoy_2023,Murguia-Berthier_2024}.

Second, our thermodynamic treatment remains simplified.
Although the simulations include MHD turbulence and rapid thermal relaxation, they do not incorporate full radiative transfer or detailed heating and cooling processes.
In reality, the thermodynamic structure of the inner disk and magnetosphere can be influenced by X-ray irradiation, magnetic reconnection, shocks, and wave dissipation \citep[e.g.,][]{Hartmann_2016}.
The simulations also adopt a relatively thick PPD with \(h_0=0.1\), larger than the aspect ratios commonly inferred for many inner protoplanetary disks \citep[e.g.,][]{dullemond2010,Bitsch_2015}. This choice was made to maintain adequate vertical resolution in the global 3-D MHD calculations.
Because the thermodynamic state determines how efficiently gas can cool and settle into a rotationally supported circumplanetary structure \citep[see, e.g.,][]{Fung_2019,Ginzburg_2019,Kuwahara_2026}, while the disk aspect ratio affects gap opening \citep[e.g.,][]{Crida_2006} and the transition between 3-D and 2-D accretion regimes, more realistic thermodynamics and aspect ratios may modify the morphology, gas supply, and long-term stability of the CPDs identified here.

Third, the planets are represented as softened gravitational potentials without explicit planetary surfaces or sub-grid accretion prescriptions such as sink cells \citep[see][]{Choksi_2023}.
This can affect the innermost circumplanetary flow and the measured angular-momentum budget \citep[e.g.,][]{Schulik_2020}.
Future work should explore improved sink prescriptions or more realistic planetary boundary conditions \citep[e.g.,][]{Kley_1999}.

Fourth, the planets are held on fixed circular orbits and are not allowed to migrate. Combined with the relatively short integration times (tens of orbits or less), this prevents us from studying the long-term coupling between migration, turbulence, and circumplanetary evolution. In strongly turbulent MHD disks, stochastic torques and time-variable magnetic stresses may substantially alter the classical Type I/II migration picture \citep[e.g.,][]{Wu_2024}. Longer-duration global 3-D MHD simulations with freely migrating planets will ultimately be required to determine whether the circumplanetary structures found here can persist over secular timescales and to quantify how gap opening, migration, and resonance capture/escape processes \citep[see, e.g.,][]{Chen_2025} modify their gas supply and angular-momentum evolution.

\section{Conclusions}\label{sec:conclusion}

In this work we extend the previous Paper I, by going into detail on the vicinity of planets.
We have run global 3-D ideal-MHD simulations of magnetospheric accretion  to study how circumplanetary disks (CPDs) form around massive planets close to the truncation radius (\(R_T\)), which is the inner edge of a protoplanetary disk (PPD).
Starting from a quasi-steady, turbulent magnetospheric accretion flow, we inserted three planets with the same planet-to-star mass ratio, $q=0.01$ ($\simeq 10\,M_J$ for a solar-mass star), at orbital radii $r_p=1.8\,R_T$, $0.9\,R_T$, and $0.45\,R_T$.
From outermost to innermost, these locations probe the MRI-active inner PPD, the disk--cavity transition region, and the low-density magnetospheric cavity.

These three planets produce three different outcomes: a coplanar CPD, a polar CPD, and no persistent structure.
The \textit{outer planet}, embedded in the inner disk, develops a stable coplanar CPD whose angular momentum is dominated by the vertical component \(L_z\), aligned with the angular momentum of the parent PPD.
The \textit{middle planet}, close to \(R_T\), forms a persistent polar CPD whose angular momentum is instead dominated by the in-plane component \(L_y\), indicating that the net rotation axis of the circumplanetary gas lies nearly parallel to the PPD midplane.
The \textit{inner planet}, located inside the cavity, does not sustain a coherent angular-momentum direction, and only sporadic envelope-like concentrations appear.
Figure \ref{fig:cpd_diagram} shows a schematic summary of these circumplanetary environments.

The \textit{outer planet} exhibits a rotationally supported CPD undergoing meridional circulation while maintaining a connection to the surrounding PPD.
In the planet-centered frame, the CPD appears as a dense, long-lived, flared equatorial structure occupying the inner \(\sim 0.3\)--\(0.4\,R_H\), while the gas rotates prograde with an azimuthal velocity that follows the local planet-centered Keplerian profile within $\lesssim 0.25\,R_H$.
It is fed from gas entering the Hill sphere from high latitudes, which is then funneled from the atmosphere towards the inner body.
A recycling pattern develops.
The gas is redirected radially outwards through the dense midplane layer, before a significant part of it is wound back and given back to the atmosphere.

While hydrodynamics dominate the large-scale circulation, magnetic stresses play a distinct role in the angular-momentum evolution for the coplanar CPD.
The magnetic energy is strongly amplified within the dense CPD body, where circumplanetary rotation winds the field into a predominantly azimuthal configuration.
Even so, the dense CPD remains mostly gas-pressure dominated (\(\beta \geq 1\)), while the surrounding low-density atmosphere is more strongly magnetized.
The radial Maxwell stress, $-\langle B_rB_\phi\rangle$, transports angular momentum outward through the disk body, thereby enabling inward accretion, while the $\theta$--$\phi$ stress transports angular momentum from the atmosphere toward the midplane.
Therefore, the meridional recycling circulation is mainly driven by hydrodynamical processes, whereas disk accretion is driven by magnetic angular-momentum transport.
The quasi-steady recycling state is maintained by nature of the radial and meridional advection terms and the stress-divergence terms largely balancing across $0.1\leq r/R_H\leq0.5$.

The polar CPD around the \textit{middle planet} shows that coherent circumplanetary disks can survive even near the magnetospheric cavity.
Its density and magnetic energy near the planet are comparable to those of the coplanar CPD, and the bound polar structure remains gas-pressure supported within its dense body.
We interpret the polar orientation as a consequence of anisotropic, magnetically controlled feeding near the truncation radius: gas entering the Hill sphere need not carry angular momentum aligned with the parent disk, while magnetic stresses may help preserve the resulting structure once established.
This orientation is not, however, a generic consequence of proximity to $R_T$, since the planet located near the truncation radius in Paper I developed a coplanar CPD.

We discuss a CPD formation criteria that provides a useful hierarchy for interpreting these outcomes.
The minimum capture condition, \(R_B/R_p>1\), is easily satisfied in the disk but becomes very restrictive in the magnetospheric cavity, due to the free-falling flow and low density.
The thermal mass parameter, \(q_{\rm th}\), provides a second diagnostic, with super-thermal planets being more likely to generate extended, disk-like circumplanetary flows, although \(q_{\rm th}>1\) is not by itself a sufficient condition for a Keplerian CPD \citep[see, e.g.,][]{Machida_2008,Fung_2019,Choksi_2023}.
In this framework, the \textit{outer planet} satisfies both the Bondi and thermal-mass criteria, the \textit{inner planet} fails the minimum capture requirement in the cavity, and the \textit{middle planet} occupies a transition regime where PPD intrusion due to a non-rotating star and dense magnetospheric filaments can make the local environment resemble the inner disk enough to sustain a CPD \citep[see][]{Zhu_2024}.

For the \textit{outer planet}, we also measured the mass inflow through the Hill-sphere surface and found that the gas supply is substantial, but implies slow late-stage growth. The one-way inflow rate reaches a broad plateau of \(\dot{M}_{p,\rm in}\simeq 7\dot{M}_{\rm Hill,2D}\), close to the sink-less hydrodynamic scaling of \citet{Choksi_2023}. Because this quantity counts all gas entering the Hill sphere, it is interpreted as an upper limit to the true planetary accretion rate, especially since much of the inflowing material participates in recycling. The resulting mass-doubling timescale is \(t_{\rm double}\gtrsim 10^7\) yr, with similar values obtained when the scaling is extrapolated to other super-thermal planets. This suggests that close-in giant planets embedded in MRI-active, magnetospherically influenced disks may experience only limited additional growth during this late stage.

The qualitative picture that emerges is that close-in massive planets can sustain CPDs in the MRI-active inner disk undergoing magnetospheric accretion.
Their morphology and dynamics are not determined by planet mass alone, but also by the magnetic, thermodynamic, and kinematic structure of the local environment. Through this environmental influence, magnetospheric accretion can shape circumplanetary structures near the inner edge of protoplanetary disks, with potential implications for satellite formation and planetary spin and mass evolution.
Whether this environmental imprint persists on secular timescales remains an open question, requiring longer simulations spanning a broader range of disk, stellar-field, and planetary parameters than the restricted set explored here.

\begin{acknowledgements}
    The simulations were carried out using NASA's Pleiades
    supercomputer. A. C. S. and Z. Z. acknowledge support from
    NASA award 80NSSC22K1413, NSF award 2408207, and NSF EPSCoR
    Research Fellows award 2429732. This work was also supported
    by NASA TCAN grant 80NSSC21K0496 and by the Nevada Center
    for Astrophysics through NASA award 80NSSC23M0104.
    This research was supported in part by grant NSF PHY-2309135
    to the Kavli Institute for Theoretical Physics (KITP).
\end{acknowledgements}

\bibliography{bb.bib}{}
\bibliographystyle{aasjournal}

\end{document}